\documentclass[11pt]{article}
\usepackage[letterpaper,margin=1in]{geometry}
\usepackage{amsmath,amssymb}
\usepackage{graphicx}
\usepackage{subcaption}
\usepackage{booktabs}
\usepackage{float}
\usepackage{xcolor}
\usepackage[numbers,sort&compress]{natbib}
\usepackage{authblk}
\usepackage[colorlinks=true,allcolors=blue]{hyperref}
\graphicspath{{images/}}

\newcommand{\degN}{^{\circ}\mathrm{N}}
\newcommand{\degS}{^{\circ}\mathrm{S}}
\newcommand{\Tm}{T_{2\mathrm{m}}}
\newcommand{\Topt}{T_{\mathrm{opt}}}
\newcommand{\Ttgt}{T_{\mathrm{tgt}}}

\newcommand{\dd}{\mathrm{d}}

\title{Optimizing Geoengineering Interventions Using Differentiable Climate Models}

\author[a]{Pulkit K.\ Dubey}
\author[b]{Dorian S.\ Abbot}
\author[a,*]{Ashesh Chattopadhyay}
\affil[a]{Department of Applied Mathematics, University of California, Santa Cruz, CA 95064}
\affil[b]{Department of the Geophysical Sciences, University of Chicago, Chicago, IL 60637}
\affil[*]{Corresponding author. E-mail: aschatto@ucsc.edu}
\date{}

\begin{document}
\maketitle

\begin{abstract}
The deployment of a geoengineering program to cool Earth's climate may be
imminent. It is crucial that tools be developed to ensure that such a program
would achieve its objectives while minimizing disruption. Here we exploit
recently developed differentiable atmospheric models to demonstrate a novel
geoengineering control strategy. In the differentiable primitive-equation
atmospheric model JAX-GCM we impose a uniform $+4$\,K ocean warming and ask what
pattern of sea-surface temperature cooling -- in five ocean-masked zonal bands
of prescribed SST forcings whose amplitudes are free -- returns land
near-surface air temperature closest to the model's own unwarmed climatology.
This idealized set-up represents a cooling pattern that could be delivered
physically either by marine cloud brightening or stratospheric aerosol
injection. Gradients through chaotic dynamics decorrelate from the true
sensitivity beyond the Lyapunov horizon, so we optimize greedily over segments
of 8 to 14 days, following receding-horizon control. The learned strategy
removes $92.3 \pm 0.4\%$ of the realized land warming across a ten-member
ensemble of two-year rollouts, and a three-year run sustains it. If we use the
spatial pattern of land temperature as the optimization objective, the
distributions of precipitation, evaporation, and specific humidity over land are
restored as well, even though they are not included in the objective function.
The learned strategy from JAX-GCM replayed in the AI emulators LUCIE and
NeuralGCM without re-optimization is successful, suggesting robustness. These
promising results demonstrate a strategy for designing optimal climate
interventions that can be applied broadly for geoengineering scenarios under
consideration.
\end{abstract}


\section{Introduction}\label{sec:intro}

Proposals to offset greenhouse warming by engineering the climate system,
e.g., marine cloud brightening \citep{latham2012mcb} and stratospheric aerosol
injection \citep{crutzen2006albedo}, pose an inverse design problem: which spatiotemporal
pattern of forcing best restores a target climate? The inverse problem
has been treated as an optimization problem since the work of
Ban-Weiss and Caldeira \citep{banweiss2010optimization}. The existing
solution families
superpose precomputed linear response patterns
\citep{banweiss2010optimization, macmartin2013tradeoffs,
brody2025optimization}, tune feedback gains on a few zonal moments
\citep{kravitz2014explicit, macmartin2014closedloop,
kravitz2017objectives}, invert Green's-function response operators
\citep{lu2020neutral, ren2024optimal}, or search by
sampling with reinforcement learning \citep{dewitt2019sai, quan2025rl} (Section~\ref{si:prior}).
Each assumes a linear response or requires an additional simulation for every added forcing parameter.

Gradient-based optimization yields sensitivities to a large number of
forcing parameters in a single backpropagation pass, circumventing the
simulation expense. Such methods are established in geophysical
modeling. Variational data assimilation minimizes forecast error by
adjoint descent over a window of hours to days
\citep{talagrand1987adjoint, rabier2000ecmwf}. The ECCO state estimate
extends the window to decades and admits the surface forcing fields as
control variables, in a coarse, non-eddying ocean where the adjoint
remains stable \citep{forget2015ecco}. In chaotic regimes the adjoint
grows exponentially with time, and the usable optimization window
shrinks to days or weeks. Conditional nonlinear optimal perturbation
operates within that horizon, optimizing an initial or boundary
perturbation for maximum error growth \citep{mu2003cnop, song2026cnop}.
Differentiable climate models supply these gradients by construction
\citep{gelbrecht2023differentiable, kochkov2024neuralgcm} and have
been used to produce worst-case hurricanes \citep{plotkin2019maximizing} and heatwaves
\citep{vonich2024predictability, whittaker2025storylines} and global
emissions trajectories \citep{womack2026scenario}. Each of these
optimizes toward observed history or maximum error growth, and the ECCO
forcing fields are constrained to stay near the observed reanalysis
forcing rather than designed (Section~\ref{si:prior}). To our knowledge, no
published work uses these gradients to design a spatiotemporal forcing
that steers a climate simulation to a prescribed counterfactual target
and holds it there over multiple years. That is the problem addressed
here.

\begin{figure*}[t!]
\centering
\includegraphics[width=\textwidth]{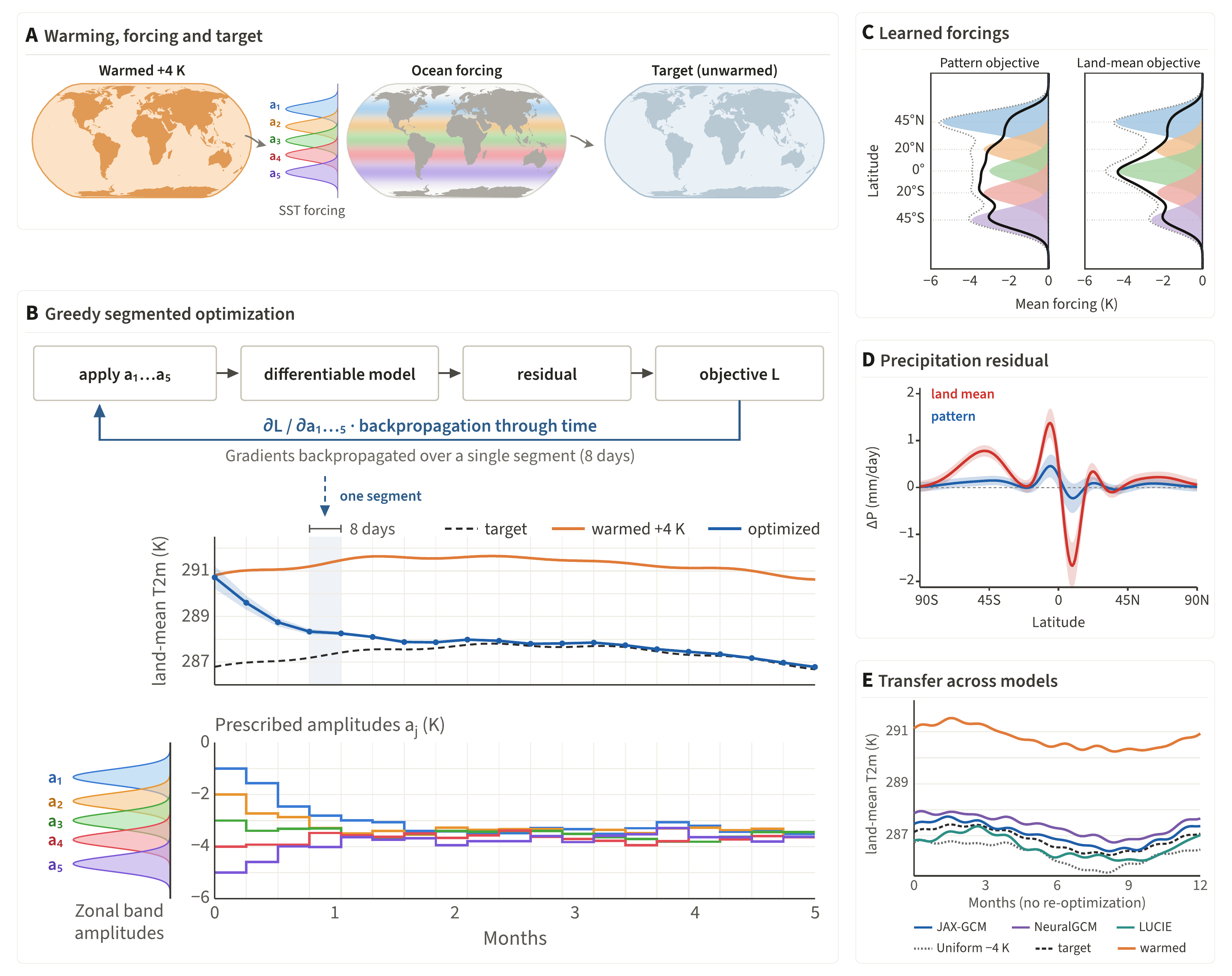}
\caption{Problem setup, optimization scheme, and outcomes.
Gradient-based optimization through a differentiable climate model
learns a time-varying pattern of ocean cooling that returns a warmed
climate to its unwarmed target. (A) A uniform $+4$\,K SST offset warms
the climate; five ocean-masked zonal bands with fixed shapes and free
amplitudes $a_1,\dots,a_5$ supply the corrective SST forcing; the
target is the model's own unwarmed climatology. (B) Greedy segmented
optimization. Amplitudes for each 8-day segment are found by
backpropagation through time, applied, and the state advances to the
next segment. No gradient crosses a segment boundary, yet land-mean
$\Tm$ converges to the target in ten weeks and holds it. (C) Time-mean
forcing learned under the pattern and land-mean objectives. (D)
Zonal-mean precipitation residual for the two objectives; the pattern
objective restores precipitation, which never enters the loss. (E) The
learned strategy, replayed without re-optimization in LUCIE and
NeuralGCM, tracks each model's unwarmed target.}
\label{fig:schematic}
\end{figure*}

Differentiability alone, however, does not make the problem tractable. A gradient
backpropagated through chaotic dynamics grows as $e^{\Lambda T}$, where
$\Lambda$ is the leading Lyapunov exponent, even as the true sensitivity of a
time-averaged objective remains bounded
\citep{lea2000sensitivity, metz2021gradients}. Beyond a few multiples of the Lyapunov horizon
$1/\Lambda$ (about one week for the mid-latitude atmosphere) a
single-trajectory gradient carries no usable signal. 
We therefore design a greedy optimizer (Fig.~\ref{fig:schematic}). The
forcing amplitudes are optimized over one short segment by exact
backpropagation through time, the converged forcing is applied, the
state advances, and the loop repeats. No gradient crosses a segment boundary 
or extends beyond two weeks, yet the concatenated strategy steers a perturbed 
climate to its target and sustains it over the following years 
(Sec.~\ref{sec:control}). This scheme is called the
receding-horizon approach of adjoint-based turbulence control
\citep{bewley2001predictive}. The
gradient provides the sensitivity to every forcing parameter in one
backward pass, at a cost independent of the parameter count, without a precomputed 
response library and without assuming that the responses to the individual bands 
superpose linearly.

To demonstrate our optimizer's success on designing strategies for climate intervention, we use JAX-GCM~\citep{davenport2026jcm}, a differentiable physics-based atmospheric model which couples the spectral
primitive-equation core of NeuralGCM \citep{kochkov2024neuralgcm} to
SPEEDY subgrid-scale physics \citep{molteni2003speedy}. We impose a uniform $+4$\,K sea surface temperature (SST) warming in JAX-GCM and ask
the optimizer to design the time-varying amplitudes of five forcings shaped as Gaussian zonal bands of perturbation to SST, with
the objective of returning near-surface air temperature over land to the
model's own unwarmed climatology. We run the same optimization in two
AI emulators, LUCIE~\citep{guan2024lucie,guan2025lucie}, built on a spherical Fourier neural
operator, and NeuralGCM, a hybrid model with physics-based dynamics and
learned subgrid-scale physics \citep{kochkov2024neuralgcm}. This setup isolates four questions that an
intervention-design pipeline must answer.

\begin{enumerate}
\item[(a)] Do backpropagated gradients reflect the true sensitivity of the objective?
\item[(b)] Does stitching short-horizon solutions produce long-horizon
      control?
\item[(c)] What does the optimizer learn beyond a uniform offset, and at
      what cost to the rest of the climate?
\item[(d)] Does a \textit{strategy} (the spatio-temporal variations of the Gaussian-shaped forcings) learned in one model apply to another?
\end{enumerate}

To address (a), we compare the backpropagated gradient against central
finite differences across segment lengths, seasons, and operating
points, and we record an ensemble noise-to-signal statistic at every
segment of every run (Section~\ref{si:grad}). To address (b) we run the greedy
trajectory over two and three years (Sec.~\ref{sec:control}). For (c)
we optimize two objectives, one on the area-weighted mean of the temperature
over land (land-mean objective)
and one on its spatial pattern (pattern objective), and measure the
response of precipitation, evaporation, and specific humidity, none of
which are used in either objective (Sec.~\ref{sec:lossdesign}). 
An objective that uses a global-mean
temperature target is known to distort the hydrological cycle
\citep{bala2008hydrological, ricke2023hydrological, lee2020expanding}.
We also compare the learned strategy against flat and linear-response
baselines at equal forcing (Sec.~\ref{sec:pareto}). For (d) we optimize
LUCIE and NeuralGCM independently and also replay the JAX-GCM strategy in each
without re-optimization (Sec.~\ref{sec:crossmodel}, Section~\ref{si:emuval}).

\section{Results}
\label{sec:results}

\subsection{Gradient-based intervention design sustains multi-year control}
\label{sec:control}

We impose a uniform $+4$\,K sea-surface temperature (SST) warming in
JAX-GCM and seek to design an intervention, an additional SST perturbation
added on top of the warming, that returns the near-surface air
temperature over land to the model's own unwarmed climatology (target).
The intervention 
consists of five zonal bands of SST perturbation
centered at $45\degN$, $20\degN$, $0^{\circ}$, $20\degS$, and
$45\degS$. Within each band the perturbation follows a Gaussian
profile in latitude of width $\sigma = 10^{\circ}$, and is applied
only over the ocean.
The value of the SST perturbation at the center of each Gaussian band
$j = 1,\dots,5$ is referred to as the amplitude $a_j(t)$ of the respective zonal 
band (Fig.~\ref{fig:schematic}A). 
Designing the intervention amounts to determining the individual band amplitudes
as a function of time in order to achieve the target. However, the
amplitudes are not adjusted continuously. 
Instead, the run is divided into consecutive 8-day blocks, which we
call segments, and the amplitudes are optimized and held constant within each
segment (Fig.~\ref{fig:schematic}B). Each $a_j(t)$ is therefore a piecewise-constant 
function of time, taking one value per segment. The optimizer selects, 
for each segment, the five amplitudes that minimize
\begin{equation}
L \;=\; \alpha\, \bigl(\langle d \rangle_w\bigr)^{2}
   \;+\; \beta\, \bigl\langle \bigl(d - \langle d \rangle_w\bigr)^{2} \bigr\rangle_w
   \;+\; \mu\, \overline{a_j^{2}}
   \;+\; \lambda\, \overline{\bigl(a_j - a_j^{-}\bigr)^{2}} ,
\label{eq:loss}
\end{equation}
where $\langle \cdot \rangle_w$ is the area-weighted
mean over land and $d(\mathbf{x})=\Topt(\mathbf{x}) - \Ttgt(\mathbf{x})$ 
is the difference between the near-surface air temperature of
the optimized run and that of the unwarmed target, each time-averaged 
over the segment. The overbar is the mean over the five bands, $a_j$ 
are the current segment's constant amplitude values, and
$a_j^{-}$ are the converged amplitudes from the previous segment. The optimizer
determines the amplitudes, evolves the model over a segment's 
duration with the selected amplitudes held constant, and
updates the amplitudes at the beginning of the following segment.

The amplitudes ($a_j(t)$) prescribed by the optimizer differ from the forcing
applied on the climate. 
A band perturbs only the region of the ocean it occupies, and each latitude
contributes in proportion to its surface area, which decreases toward
the poles as the cosine of latitude ($\cos\varphi$). 
The applied forcing profile is therefore the summed prescribed profile 
weighted by the zonal-mean ocean fraction and by $\cos\varphi$ (Eq.~\eqref{eq:forcingprofile}).
Band $j$'s contribution to the global-mean applied SST perturbation is thus 
given by $F_j(t) = a_j(t) A_j$, where $A_j$ is the band's effective ocean area as a 
fraction of global area (Eq.~\eqref{eq:bandforcing}, Section~\ref{si:areaweight}). The five amplitude histories 
together with the fixed spatial pattern of the bands yield the spatio-temporal 
variation of the applied forcing, which we call the strategy.

The squared land-mean bias and the spatial variance of the residual, $d$, 
measure the distance of the optimized state from the target. 
For $\alpha = \beta = 1$ they sum to the
land-mean squared error $\langle d^{2} \rangle_w$ (Section~\ref{si:lossdecomp}). 
The land-mean objective ($\beta = 0$) retains only the bias term, 
while the pattern objective ($\alpha = 1$, $\beta = 0.5$) penalizes the spatial
structure of the residual as well. The last two terms regularize the
intervention, penalizing large amplitudes ($\mu$)
and their change between segments ($\lambda$). The segments are optimized 
in succession over multi-year runs (Materials and Methods). Fig.~\ref{fig:schematic} summarizes the method.

The following conventions are followed unless stated otherwise. The unwarmed control and the uncontrolled
$+4$\,K reference are five-member ensemble means. Pattern-objective results
are averages over ten ensemble members that differ in their optimizer
initialization, land-mean-objective over five, and shading or error bars
span $\pm 1$ standard deviation across members. All statistics are time means over the final
364 days, one seasonal cycle, of runs two years or longer. 

\begin{figure*}[t!]
\centering
\includegraphics[width=\textwidth]{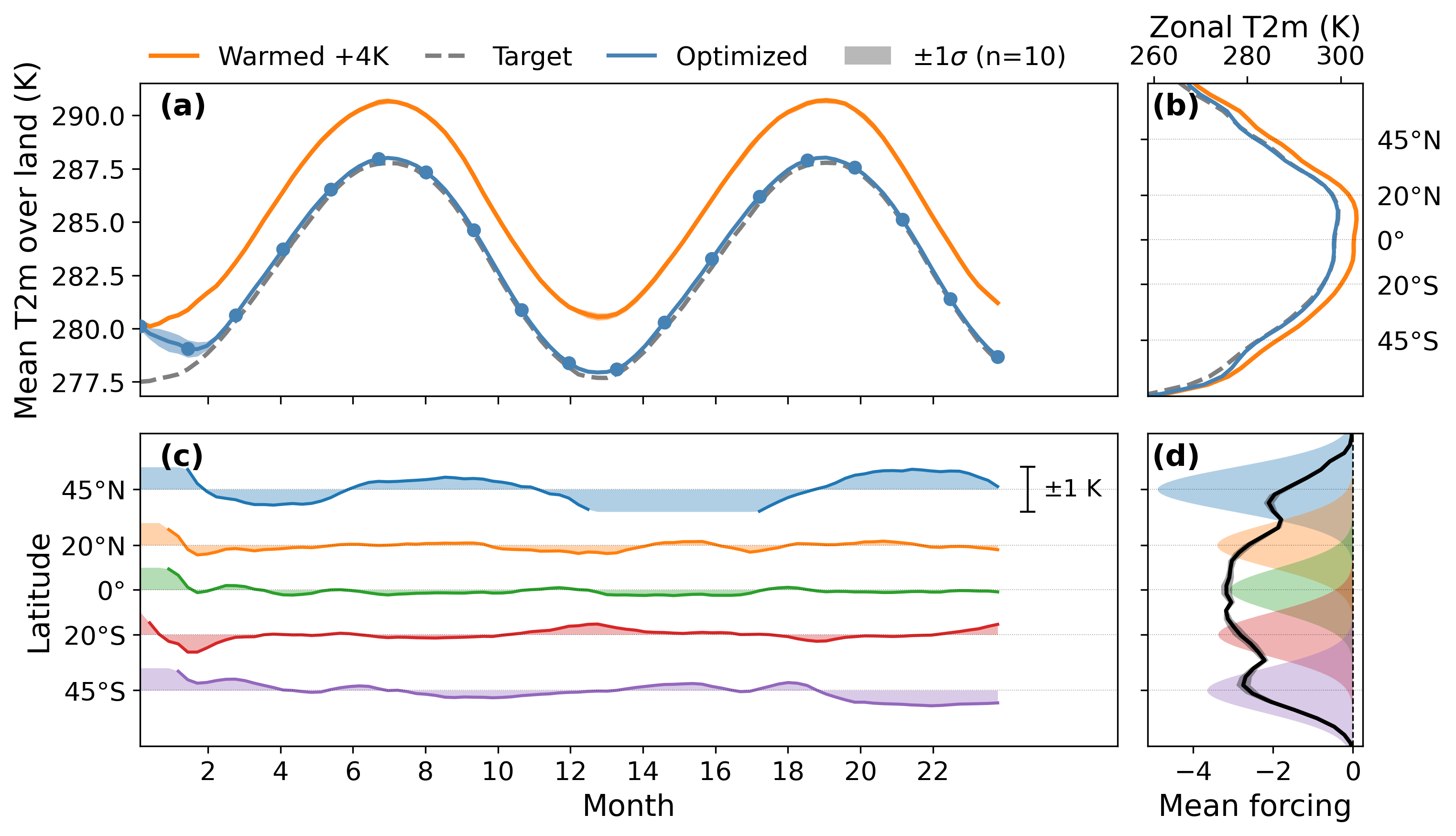}
\caption{Segment-wise control with the temperature pattern based objective 
via sea-surface-temperature (SST) forcing in JAX-GCM. 
Each segment (optimization window) is 8 days, but the 
optimizer reduces the mean temperature to the target over 10 weeks and 
sustains it over two years. All time series plots are shown as a moving average
over 5 segments (40 days).
(a)~2m temperature averaged over all land areas, $\Tm$ (K), vs time (months) 
for the warmed ($+4$\,K), target, and optimized trajectories. The optimized 
curve shows every fifth segment with a dot. 
(b)~Zonal-mean $\Tm$ (K)
against latitude. The optimized zonal mean is close to that of the target 
(unwarmed) state.
(c)~Temporal variations in the learned strategy. Each band's cooling 
amplitude is shown as the deviation from its time-averaged value in (d), 
drawn about the band's own latitude; the bar at the upper right gives the 
scale, $\pm 1$\,K.
(d)~Time-averaged values of the applied cooling against latitude, with the average taken over the full two-year run. 
The filled curves denote the five Gaussian bands 
with the amplitudes indicated by their heights. Thick black line shows the 
ocean-fraction and area-weighted forcing profile of Eq.~\eqref{eq:forcingprofile}.}
\label{fig:longrun}
\end{figure*}

Figure~\ref{fig:longrun} shows the temperature reduction achieved with 8-day
optimization windows over a two-year run (91 segments).
The optimized run converges toward the target climatology within about
10 segments, then tracks the unwarmed seasonal
cycle for two years, removing over 90\% of the land-mean
warming realized in the uncontrolled $+4$\,K run. The land-mean
trajectories of the ensemble members differ only in the initial 10 weeks
and converge beyond it. The ensemble members are generated by varying
the initial band amplitudes. Ensembles generated by varying the initial
condition instead show a statistically indistinguishable spread (Section~\ref{si:controlstats}). The optimized zonal-mean near-surface temperature profile
follows the target at all latitudes (Fig.~\ref{fig:longrun}(b)).
The outcome does not depend on the optimization horizon up to a limit
(Section~\ref{si:grad}). A 14-day-segment ensemble over three years (Fig.~\ref{fig:control14})
converges to a similar spatiotemporal variation of band amplitudes and
removes a comparable fraction of the warming (Section~\ref{si:controlstats}).

The accuracy and precision of the 
backpropagated gradients, which underlie the optimization, were measured using 
finite-difference tests and noise-to-signal ratios, respectively. Both degrade
beyond a 14-day horizon and fail by about 28 to 30 days (Section~\ref{si:grad}).

The time-mean strategy the optimization produces is shown in
Fig.~\ref{fig:longrun}(d). Each filled Gaussian profile denotes a zonal
band's latitudinal extent, with its height indicating the band's
time-averaged prescribed amplitude. The thick black line is the
corresponding applied forcing profile (Eq.~\eqref{eq:forcingprofile}). The prescribed
amplitude and the applied forcing differ most for the $45\degN$ band,
which has the highest prescribed amplitude but the smallest applied
forcing, as it acts on the smallest ocean area.

\subsection{A temperature pattern objective restores precipitation and evaporation as well}
\label{sec:lossdesign}

Both objectives of Eq.~\ref{eq:loss} act only on near-surface temperature
over land. The land-mean objective ($\alpha = 1$, $\beta = 0$)
penalizes only the difference between the
segment-averaged land-mean temperature of the optimized run and that of
the target, thus targeting only for the spatiotemporal average of
near-surface temperature over land return to the target value. The
pattern objective ($\alpha = 1$, $\beta = 0.5$) is a mean-squared error
modified to weigh the land-mean bias and the spatial structure of the
land temperature residual separately, the bias at full weight and the spatial
variance of the residual at half weight (Section~\ref{si:lossdecomp}).  
Results are robust to minor changes in their relative weights.

\begin{figure*}[t!]
\centering
\begin{subfigure}[b]{0.22\textwidth}
\centering
\includegraphics[width=0.714\linewidth]{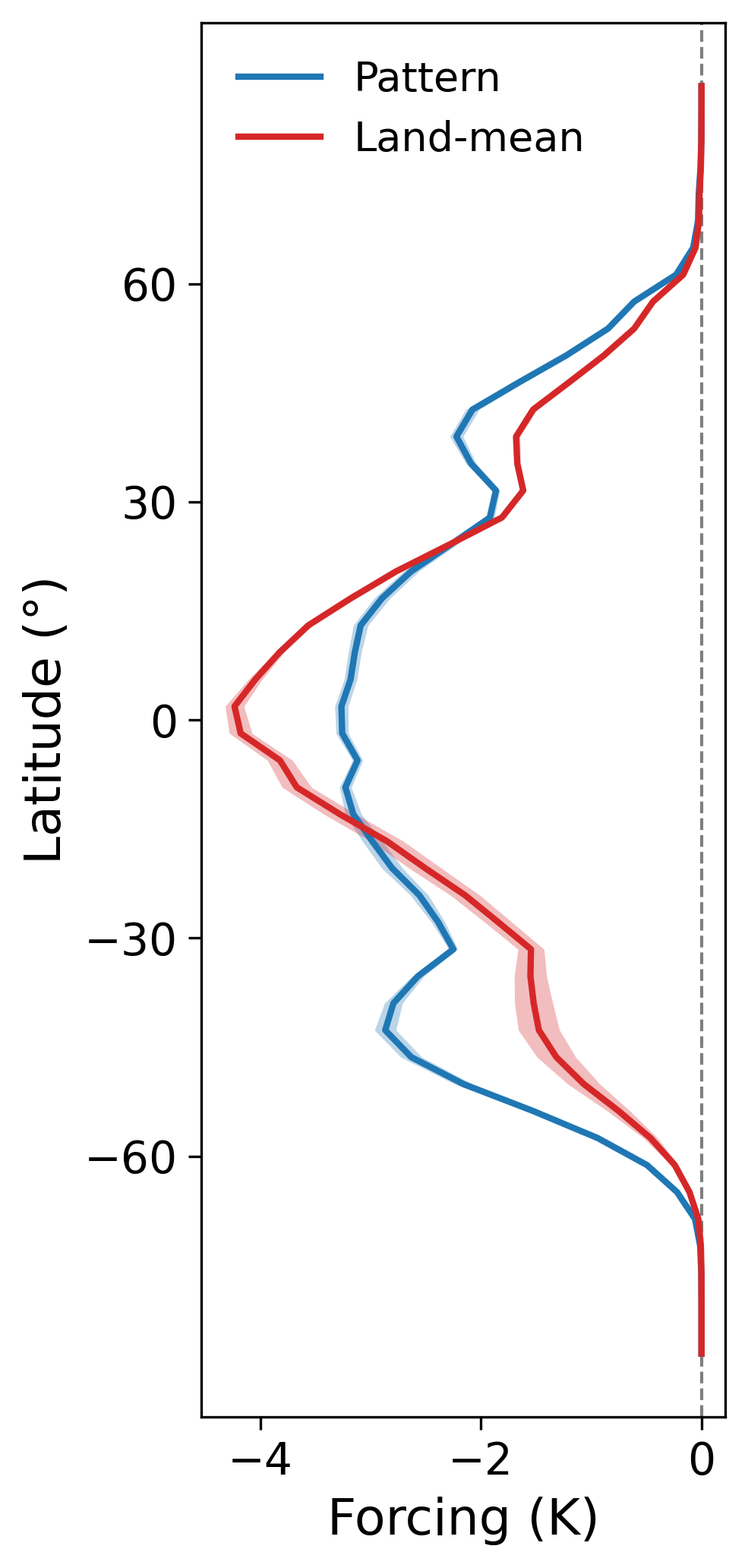}
\caption{Applied forcing}
\label{fig:lossdesign-forcing}
\end{subfigure}%
\hfill
\begin{subfigure}[b]{0.76\textwidth}
\centering
\includegraphics[width=\linewidth]{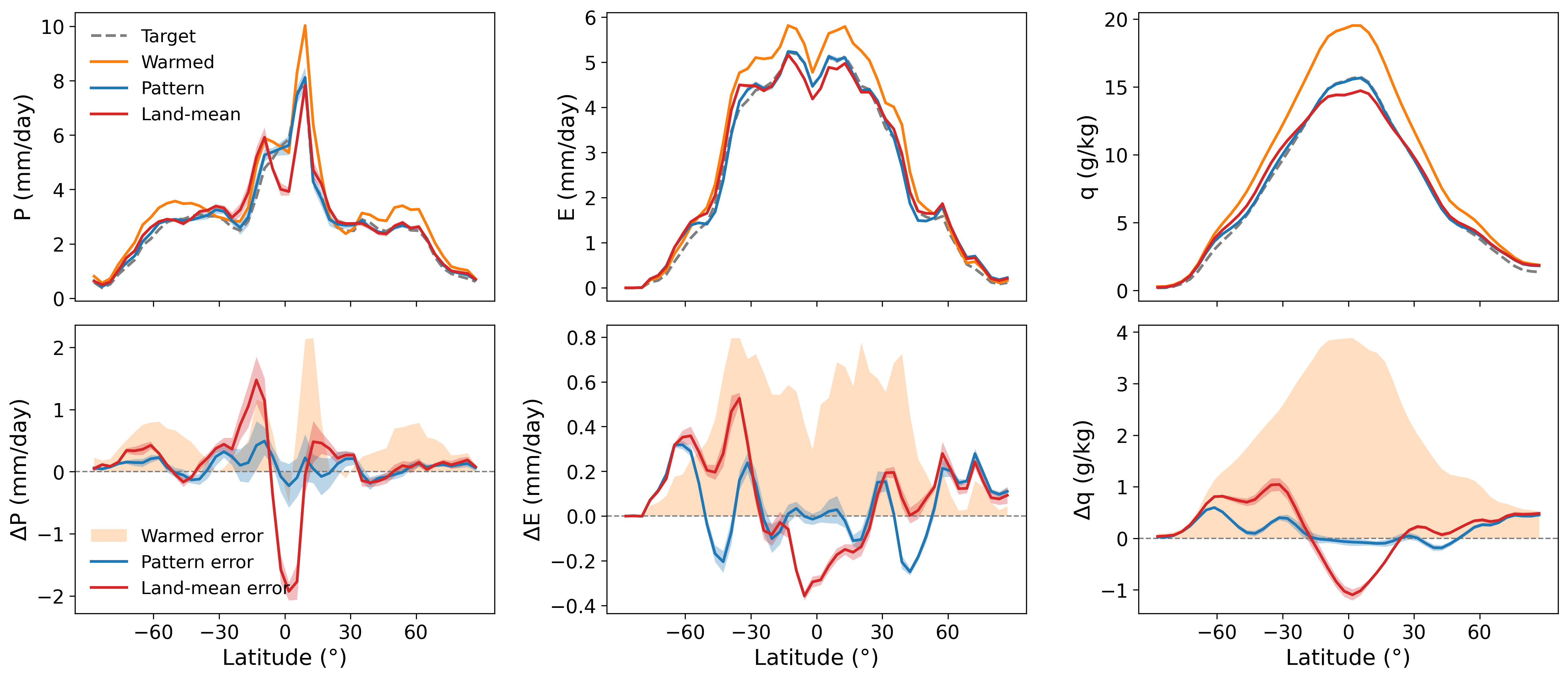}
\caption{Hydrological response}
\label{fig:lossdesign-hydro}
\end{subfigure}
\caption{Applied forcing and zonal-mean hydrological response under the two land-temperature objectives. The hydrological variables are out-of-loss (not explicitly restored by the optimization scheme).
The pattern objective applies less cooling at the equator and restores the hydrological variables better than the
land-mean objective. (a)~Ocean-fraction and area-weighted forcing profile of Eq.~\eqref{eq:forcingprofile} (K) against latitude, time-averaged over the final 364 days, for the runs optimized against the temperature pattern (Pattern, blue) and against the land-mean temperature (Land-mean, red). (b)~Top row: zonal-mean profiles of precipitation
(mm\,day$^{-1}$), evaporation (mm\,day$^{-1}$), and specific humidity
(g\,kg$^{-1}$) against latitude for the unwarmed target (dashed), the
uncontrolled $+4$\,K warmed system (orange), and the same two optimized runs. Bottom row: departures of the warmed and optimized systems
from the unwarmed target, with the $+4$\,K departure drawn as the shaded
envelope. The land-mean objective leads to a drying of the equatorial region, whereas the
pattern objective keeps residuals near zero.}
\label{fig:lossdesign}
\end{figure*}

While both strategies remove a 
comparable fraction of the realized land-mean warming (Section~\ref{si:controlstats}),
they achieve it via different strategies. The land-mean optimizer
concentrates its cooling on the equator (Fig.~\ref{fig:lossdesign-forcing}). Its $0^{\circ}$ band takes the
deepest time-mean amplitude and carries 34\% of the total applied
forcing (Fig.~\ref{fig:forcing-scalar}, Eq.~\eqref{eq:bandforcing}). The pattern optimizer
applies shallower cooling at the equator and spreads the rest across the bands (Fig.~\ref{fig:lossdesign-forcing}),
with an effort about 10\% larger than the land-mean strategy's
(Eq.~\ref{eq:effort}). The two strategies thus reach the same land-mean
temperature while placing the forcing at different latitudes.

The consequences of the differences in forcing structure between the two objectives
extend to variables unconstrained by the objective (loss) function. Out-of-loss 
variables test the extent to which a strategy restores the full climate as opposed to merely 
transferring residuals to unconstrained fields. The in-loss variable 
in both objectives is only the near-surface temperature over land, 
averaged in different ways. Since 
neither objective constrains the hydrological cycle, we evaluate 
precipitation, evaporation, and specific humidity as out-of-loss 
variables (Fig.~\ref{fig:lossdesign-hydro}, absolute profiles in the top row, 
departures from the unwarmed target in the bottom row).

Land-mean based control leaves its largest residuals on the equator. It
replaces the warmed run's excess tropical precipitation with an
equatorial deficit of comparable magnitude and opposite sign, with similar
effects on specific humidity and evaporation 
(Fig.~\ref{fig:lossdesign-hydro}, bottom row). This deficit may be attributed to
the heavy equatorial forcing under the land-mean objective (Fig.~\ref{fig:lossdesign-forcing}). The
$0^{\circ}$ band lies under the Intertropical Convergence Zone (ITCZ),
and the strong time-mean cooling applied there suppresses tropical
convection and dries the ITCZ. Because the objective sees only the
land-mean temperature, the intervention removes the warming while
disrupting the hydrological cycle. Pattern temperature control, by contrast, 
drives the zonal-mean residuals of all three hydrological variables being near zero, with an equatorial precipitation residual almost an order of magnitude
smaller than under the land-mean temperature objective.

\subsection{Optimization performance compared with baselines}
\label{sec:pareto}

We compare the performance of the learned strategies using two metrics --
the restoration gain $G$, which is the factor by which an intervention 
reduces the mean-square error of the time-averaged fields relative to 
the uncontrolled $+4$\,K run, and the effort $E$ (in K) which is the overall 
forcing prescribed by the intervention, measured as the spatiotemporal 
average of the magnitude of the forcing summed over all bands. 
A good strategy should deliver high gain at low effort. 

\begin{figure*}[t!]
\centering
\includegraphics[width=\textwidth]{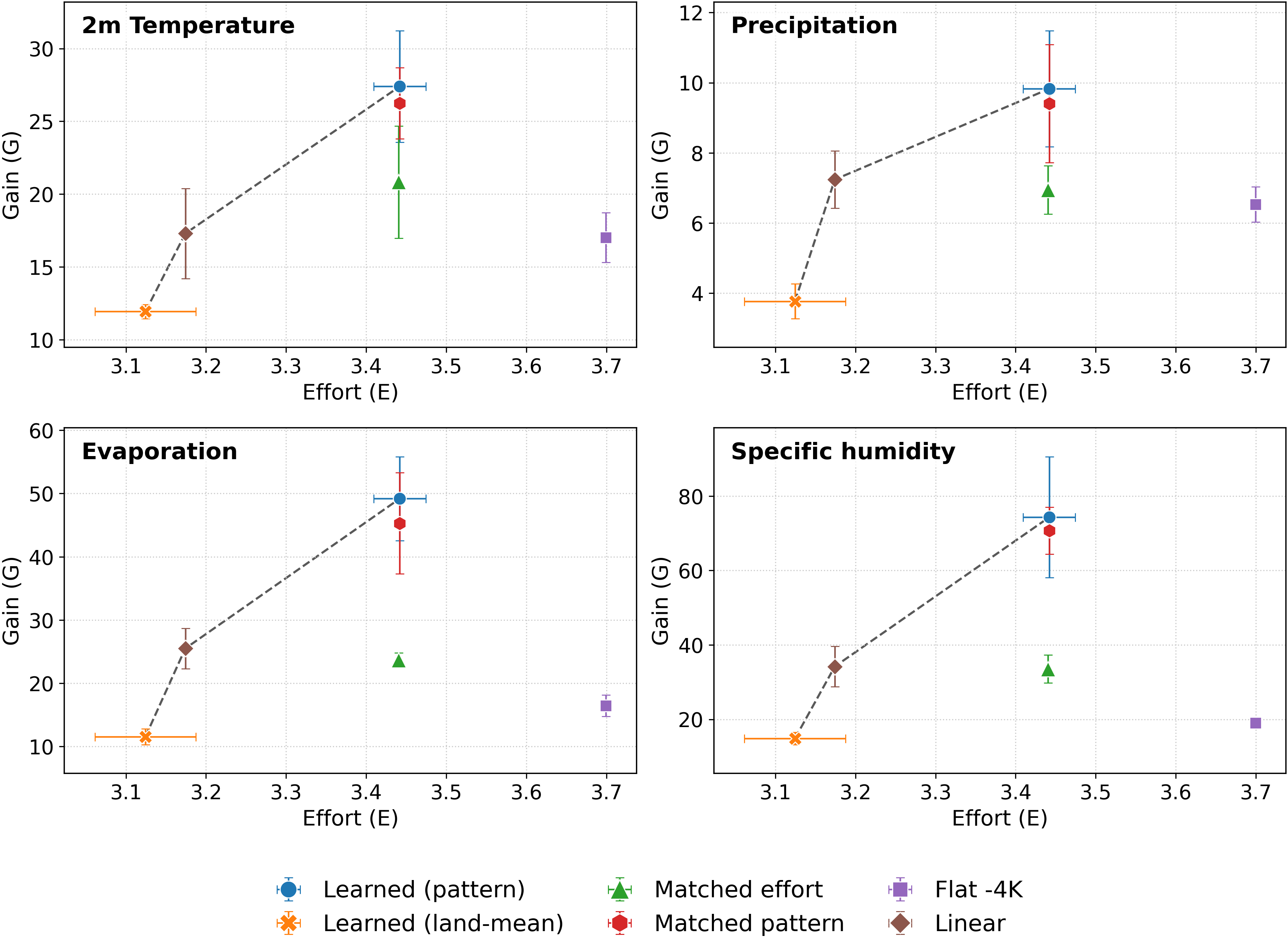}
\caption{Skill and effort of the learned strategy against baselines.
The learned strategy and its constant-in-time matched-pattern baseline
reach the highest gain on every variable and beat the flat forcing
matched to the same prescribed effort, so the spatial structure of the
forcing contributes beyond the amount of cooling. 
Each panel shows one variable, with the restoration gain $G$ of
\eqref{eq:gain} against the prescribed effort $E$ of \eqref{eq:effort}.
$G$ is the ratio of the area-averaged mean-square error of the $+4$\,K warmed run
to that of the optimized run, both measured on time-mean
fields over land against the unwarmed target.
$G = 10$, for instance, indicates that the error between the optimized and 
target runs is one tenth of that between the $+4$\,K warmed and 
the same target runs. $G = 1$ indicates no improvement.
The effort $E$ is the global-area and time mean of
the magnitude of the summed prescribed cooling profile, in K. 
Learned (pattern) and Learned
(land-mean) are the optimized strategies under the two objectives.
Matched effort applies one constant amplitude on all bands with the
effort of Learned (pattern). Matched pattern applies the time-mean
amplitudes of Learned (pattern). Flat $-4$\,K applies $-4$\,K
on every band. Linear solves for constant amplitudes under a
linear-response assumption and runs the full model with solved amplitudes. 
The dashed line joins the strategies with the largest gain for a given effort.}
\label{fig:pareto}
\end{figure*}

Fig.~\ref{fig:pareto} compares the two learned strategies against four
baselines on a gain $(G)$ vs effort $(E)$ 
(\eqref{eq:effort} and~\eqref{eq:gain}, Materials and Methods) plane.
The four baselines fall into two groups, ablations of the learned
strategy (with pattern objective) and independent references. The first
reference is the uniform baseline which applies $-4$\,K on all five
bands, as a naive solution to the $4$\,K ocean warming. The second
reference is the linear baseline that solves for constant amplitudes
under the linear-response assumption of the optimal-pattern methods
\citep{banweiss2010optimization, lu2020neutral} and runs the full
nonlinear model at those amplitudes (Section~\ref{si:variants}). The two ablations
of the learned pattern strategy are created to isolate the effects of
spatial and temporal variation of the applied forcing. The
matched-effort baseline applies one constant amplitude to all five bands
($-3.72$\,K), chosen such that its effort equals the learned strategy's,
$E = 3.44$\,K. Its gain measures what the learned effort achieves
without either the spatial distribution or the time variation. The
matched-pattern baseline constructs the amplitudes as time averages of
individual amplitudes from the learned strategy with pattern objective,
so it matches the learned spatial distribution of the forcing pattern
and its effort but not its temporal variation. The gain difference
between the learned strategy and matched-effort is the contribution of
the spatial distribution. The difference between the learned strategy
and matched-pattern is the contribution of the time variation.
All baselines are evaluated as five-member ensembles and hold their band 
amplitudes constant in time.

The learned strategy with pattern objective and its constant-in-time matched-pattern baseline have the highest gain for each variable. The uniform baseline spends more effort than the pattern strategy
($E = 3.70$\,K against $3.44$\,K) and gains less on every variable,
so a larger forcing does not imply better restoration. At equal
effort the pattern strategy beats the matched-effort baseline on
2\,m temperature ($G = 27.4$ against $20.8$), evaporation ($49.2$
against $23.6$), specific humidity ($74$ against $34$), and
precipitation ($9.8 \pm 1.7$ against $6.9 \pm 0.7$). The
matched-pattern baseline reproduces the learned gains to within the
ensemble scatter, indicating that the contribution of time variation of the amplitudes (Fig.~\ref{fig:seasonality}) is not resolved by these time-mean metrics.
The land-mean strategy spends the least effort and gains the least. The linear baseline spends less effort than others (except the land-mean strategy) and is not dominated for any variable, but gains less than the learned (pattern) strategy for all variables.

\subsection{Cross-model agreement and transfer}
\label{sec:crossmodel}

\begin{figure*}[t!]
\centering
\includegraphics[width=\textwidth]{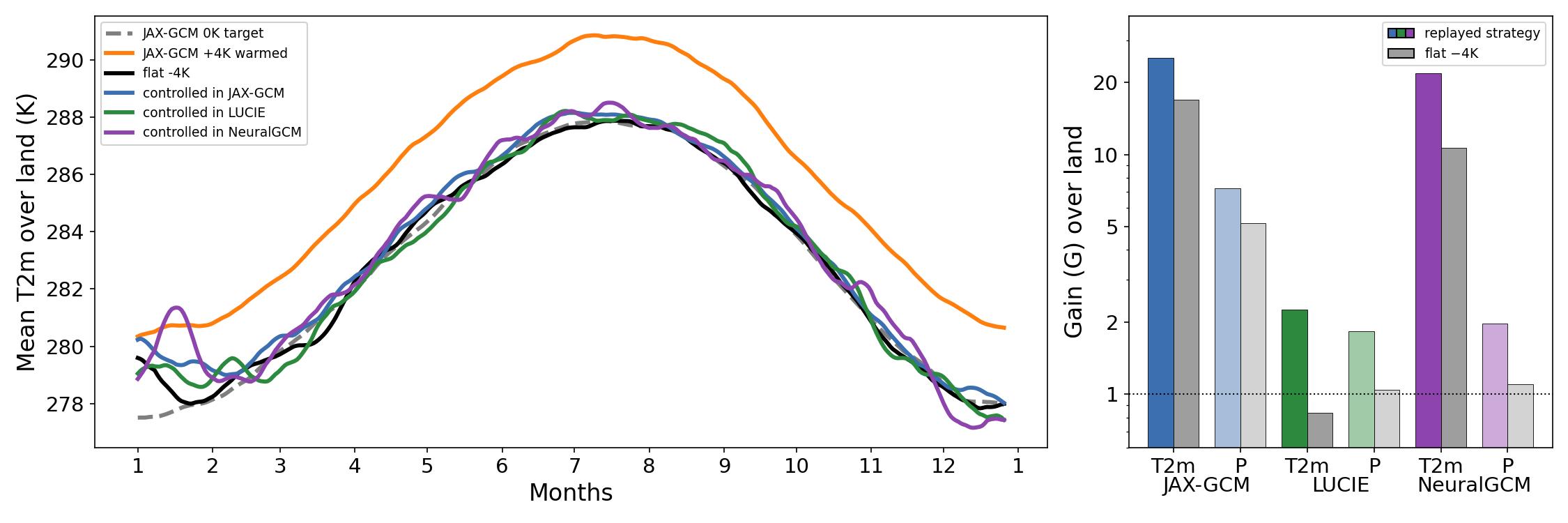}
\caption{The JAX-GCM 8-day band strategy of
Fig.~\ref{fig:longrun},
replayed calendar-aligned through all three models (JAX-GCM, LUCIE,
NeuralGCM)
running forward only, alongside a flat $-4$\,K baseline. Left: land-mean
near-surface temperature (K) against time (months) over the replay
year, with the JAX-GCM unwarmed target (grey dashed), the $+4$\,K warmed
reference (orange), the flat baseline in JAX-GCM (black), and the
controlled trajectory in
each model. Each model's trajectory is drawn as its own anomaly against
its own unwarmed control, added back onto the reference model's curve.
All time series are shown as a moving average over 15 days.
Right: restoration skill $G$ over land, the mean-square error of the
time-mean field against the model's own unwarmed control for the uncontrolled
$+4$\,K run, divided by the same quantity for the run shown.
Dark bars are the $\Tm$ field, light bars the
precipitation field. Each model carries one pair of bars
for the replayed strategy and one for its own flat $-4$\,K baseline;
the dotted line at $G=1$ is no improvement over the uncorrected
warming. In every model and both variables the strategy scores above
the baseline.}
\label{fig:replay}
\end{figure*}

The strategy learned in JAX-GCM is not specific to this model. We
replay the 8-day band strategy of Fig.~\ref{fig:longrun}
through LUCIE and NeuralGCM, each running forward only
with no re-optimization, alongside the same replay in JAX-GCM
and compare the gain $(G)$ to those via the uniform $-4$\,K baseline (Fig.~\ref{fig:replay}). Since NeuralGCM
becomes unstable beyond roughly one year,  
we restrict the runs to one-year in every model, and treat the first 60 days as transients. Fig.~\ref{fig:replay} shows the land-mean variation from the replay in different models and gain comparison between the replay and the $-4$\,K baseline. Although the flat baseline is successful in restoring the land-mean temperature, the learned strategy outperforms the flat $-4$\,K baseline on both temperature and precipitation pattern restoration in every model (Fig.~\ref{fig:replay}).
The zonal-mean profiles behind these scores, and a zonal-profile
restoration skill computed from them, are shown in Section~\ref{si:transfer}
(Fig.~\ref{fig:replay-zonal}).
Note that near-surface temperature is a different quantity on a different grid in each model, so the bar heights of Fig.~\ref{fig:replay} are computed against the model's own unwarmed control.

\begin{figure*}[t!]
\centering
\begin{subfigure}[t]{0.48\textwidth}
\centering
\includegraphics[width=\textwidth]{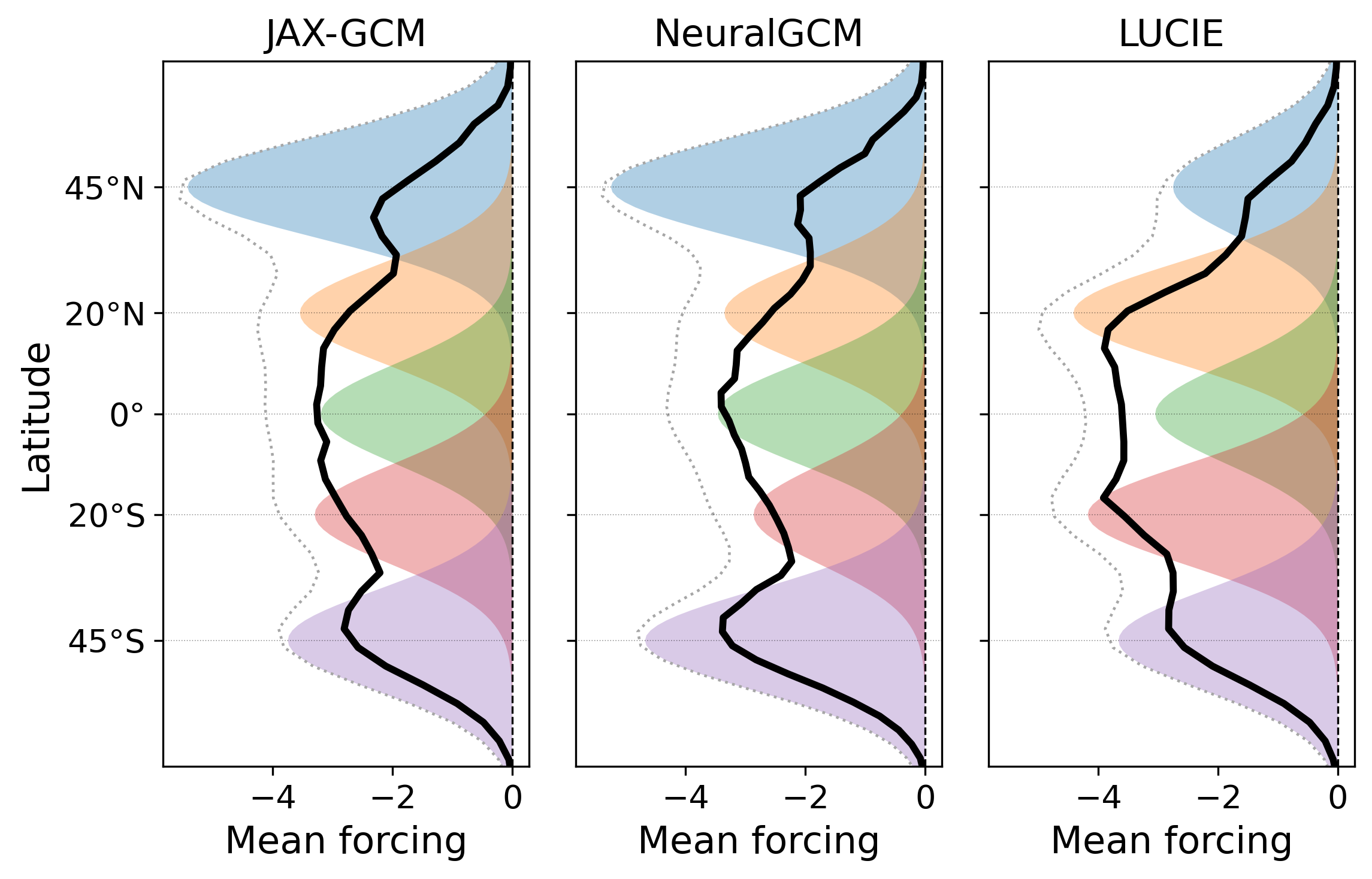}
\caption{Pattern objective}
\label{fig:forcing-pattern}
\end{subfigure}%
\hfill
\begin{subfigure}[t]{0.48\textwidth}
\centering
\includegraphics[width=\textwidth]{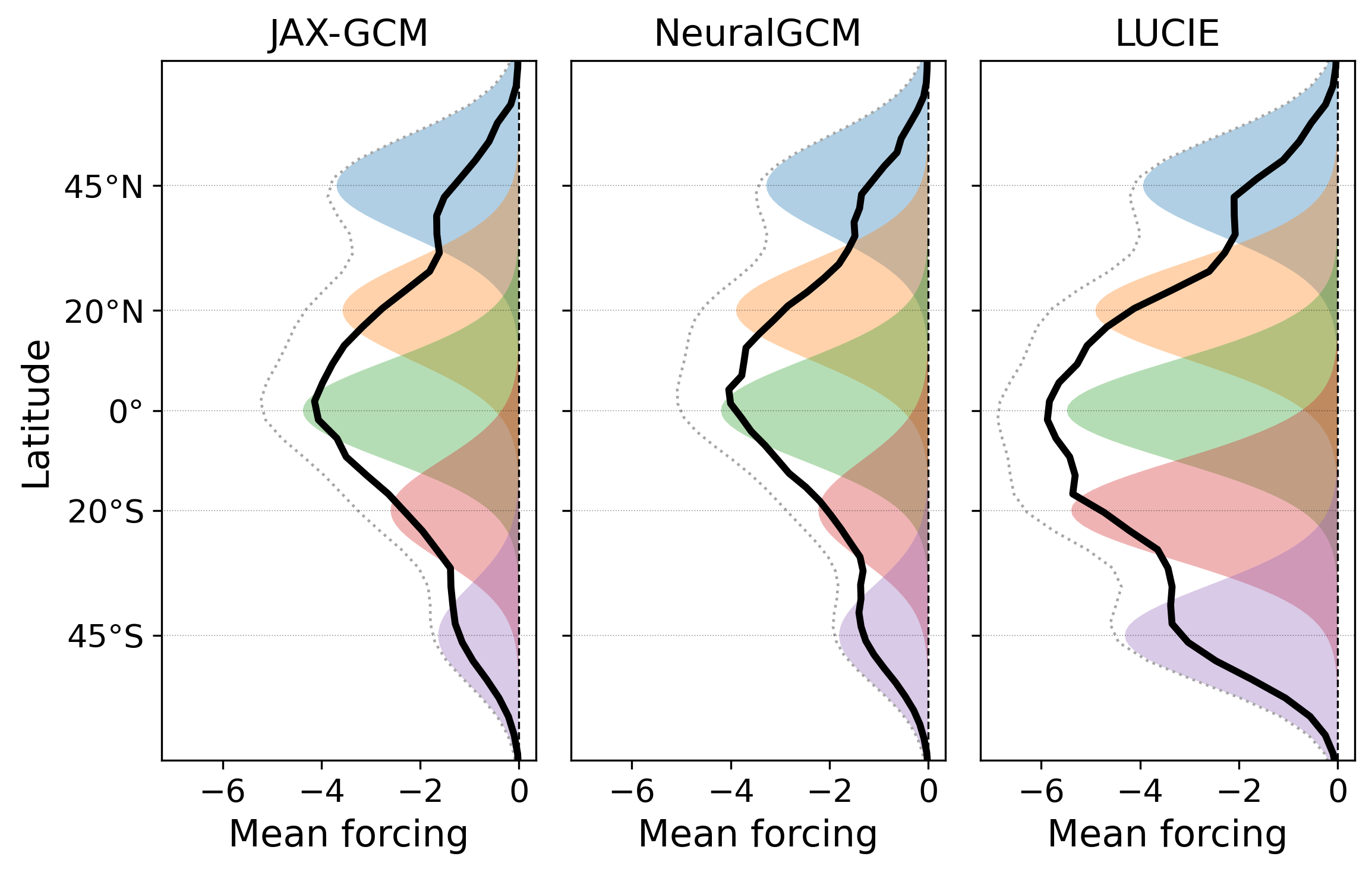}
\caption{Land-mean objective}
\label{fig:forcing-scalar}
\end{subfigure}
\caption{Time-mean forcing in the three models under (a)~the pattern
objective and (b)~the land-mean objective. Each model's subplot shows
the five Gaussian bands filled at that run's time-mean
amplitudes (K), the raw summed amplitude profile (dotted), and the
ocean-fraction- and area-weighted forcing profile of
Eq.~\eqref{eq:forcingprofile} (thick black), against latitude on a
forcing scale shared across each subfigure. Models left to right in each
group: JAX-GCM, NeuralGCM, and LUCIE. The
pattern-objective runs are the JAX-GCM run of
Fig.~\ref{fig:longrun} (two years), the run of Fig.~\ref{fig:ngcm-control}
(NeuralGCM, one year), and that of
Fig.~\ref{fig:lucie-control} (LUCIE, one year); the land-mean runs
are a two-year JAX-GCM run identical
except for its objective, a 42-segment
NeuralGCM run, and a five-year LUCIE run. In (b) all three models
concentrate the forcing on the equatorial band; in (a) the forcing is
spread, with deepest amplitude at $45\degN$ in JAX-GCM and NeuralGCM.}
\label{fig:forcing-crossmodel}
\end{figure*}

The transfer is consistent with part of what LUCIE and NeuralGCM learn
on their own. Figure~\ref{fig:forcing-crossmodel} compares the time-mean forcing each model learns under the two objectives. Under the land-mean objective all three
concentrate their forcing on the equatorial band
(Fig.~\ref{fig:forcing-scalar}), the placement that dries the ITCZ in
JAX-GCM (Fig.~\ref{fig:lossdesign-hydro}). Under the pattern
objective all three spread the forcing across the bands
(Fig.~\ref{fig:forcing-pattern}), although the spatial structure of forcing in LUCIE differs from that in the other two models. NeuralGCM converges on the structure found in JAX-GCM, applying the largest time-mean cooling on the $45\degN$ band with a secondary maximum at $45\degS$. LUCIE places its largest cooling on the subtropical bands and applies its smallest forcing at $45\degN$.

\section{Discussion}
\label{sec:discussion}

Our results demonstrate differentiable climate control as an
inverse-design method and emphasize the importance of the choice of objective. 
Interestingly, matching the spatial pattern of land temperature is enough to 
achieve good matching of hydrological variables, even though they do not appear 
in the loss function.  In feedback-regulated
multi-latitude stratospheric aerosol injection, temperature objectives
built from the global mean and two zonal moments did not by themselves
restore precipitation, which was restored only when included explicitly
as a control objective \citep{lee2020expanding}. The temperature target
here is the full land $\Tm$ pattern instead of moments, and the
forcing is applied on the surface directly rather than through a constrained
aerosol loading. The improvement of precipitation under this
temperature-only pattern objective suggests that resolving the
temperature structure associated with hydrology may provide a
parsimonious objective. A lesson from this work is that objectives for intervention design should be
evaluated on out-of-loss variables as well as on the quantities they
optimize.

Backpropagation based gradient computation scales well to higher-
dimensional problems. Optimal-pattern methods estimate
the response to each basis function separately and assume that the
responses combine linearly
\citep{banweiss2010optimization,macmartin2013tradeoffs}. Feedback
approaches regulate a small number of climate moments with a linearized
design model \citep{kravitz2014explicit,kravitz2017objectives}.
Sampling-based approaches incur search costs that grow with the action
space \citep{dewitt2019sai,quan2025rl} dimension. Reverse-mode 
differentiation, in contrast, returns the sensitivity to every forcing 
parameter in one backward pass,
without a parameter-wise response ensemble or a linearized model. As
an example of a higher dimensional forcing experiment, we consider 
a forcing space of 30 two-dimensional Gaussian SST patches (instead
of 5 zonal bands) such that their superposition yields the zonal basis 
(Section~\ref{si:variants}). The computational cost of the 30-dimensional gradient
computation remains nearly equal to that for the 5-dimensional problem.
However, the patch basis shows less restoration skill 
than the zonal basis, potentially due to a more complex optimization 
landscape in the higher-dimensional forcing space.

The learned strategies may depend on the model that provides the
gradients. Under the pattern objective LUCIE places its deepest cooling
on the subtropical bands, whereas JAX-GCM and NeuralGCM place theirs at
$45\degN$ (Fig.~\ref{fig:forcing-pattern}). A coordinated
intercomparison of AI models under the same prescribed $+4$\,K SST
warming finds that purely data-driven emulators depart from
physics-based solvers under out-of-training-distribution forcing and attributes
the departure to their treatment of land cells \citep{merchant2026zones}.
Our objectives act on land temperature under out-of-training-distribution
forcing, so LUCIE's departure is consistent with this failure mode.

The matched-pattern strategy scores nearly the same as the learned one
(Sec.~\ref{sec:pareto}), even though a clear seasonal cycle exists for
some band amplitudes (Section~\ref{si:seasonality}). In the final year the learned
amplitude vector departs from the matched constant by a seasonal cycle
with a peak-to-peak swing of $2.7$\,K in the $45\degN$ band. While the 
cycle is clear, the gain in skill over the matched-pattern
strategy is minimal because the loss is insensitive to a
displacement of this size about its minimum.
Broadly, the loss, as a function of the amplitude vector, is a
shallow parabola in amplitude space, and the optimizer brings the
amplitudes close to its bottom. The learned (seasonal) amplitude vector
differs from that of the matched-pattern by a seasonal displacement of about
$1$\,K, roughly $10\%$ of the vector's norm. Because the amplitudes sit
near a quadratic minimum, that change in the amplitude vector leads to a
much smaller change in the loss terms, about $0.007$\,K$^2$ per segment
against a weather noise of $0.5$\,K$^2$ per segment and $0.04$\,K$^2$
after ten members and a year. As a result, the seasonal cycle of the optimized amplitudes provides a very small additional gain relative to applying their time-mean values.


The moving-horizon procedure yields a feasible rather than a provably
optimal strategy. Prescribed SST cooling
omits ocean adjustment, radiative forcing, aerosol microphysics,
energy-budget closure, and deployment constraints. Hence, these results 
represent an algorithmic demonstration in an idealized inverse-design problem 
rather than a deployable intervention.

These limitations suggest future work. The objectives can be restricted
to specific areas of the land or specific time of a calendar year. 
Seasonally resolved objectives would show if the temporal variation of the
forcing contributes to the optimization. The greedy
segment-by-segment update can be replaced by model predictive control,
which optimizes several segments ahead and commits only the first. The
forcing dimensionality can be increased from five fixed bands to 
finely resolved two-dimensional patches (Section~\ref{si:variants}) or bands with widths and centers as free parameters. 
For computational efficiency, an L-BFGS optimizer
and early truncation of the per-segment iterations could reduce the 
per-segment optimization cost. These extensions would move 
differentiable climate control from a low-dimensional demonstration 
toward efficient exploration of higher-dimensional intervention design spaces 
for which existing parameter-wise response calculations become prohibitively expensive.

\section{Materials and Methods}\label{sec:methods}

\subsection*{Models}

JAX-GCM \citep{davenport2026jcm} is the spectral
primitive-equation core of NeuralGCM \citep{kochkov2024neuralgcm}
coupled to SPEEDY physics \citep{molteni2003speedy}, run at T31
resolution ($96 \times 48$) with eight vertical levels, a 30-minute
step, and prescribed SST. LUCIE~\cite{guan2024lucie,guan2025lucie} is a spherical Fourier neural operator
\citep{bonev2023sfno} trained on ERA5 \citep{hersbach2020era5,
guan2024lucie}, on a $96 \times 48$ grid with a 6-hour step. NeuralGCM
couples the same class of spectral core to learned physics
\citep{kochkov2024neuralgcm} and runs at $128 \times 64$. All three are
differentiable end to end in their native frameworks, and the same
optimization loop runs in each (Section~\ref{si:emuval}).

JAX-GCM has no 2-m diagnostics. Its $\Tm$ is a fixed blend of the skin
temperature and SPEEDY's surface air temperature $t_0$, the lowest-level
air temperature extrapolated to the surface along the local lapse rate,
$\Tm = T_{\mathrm{skin}} + 0.585\,(t_0 - T_{\mathrm{skin}})$.
LUCIE outputs 2~m temperature directly, as a variable trained on the
ERA5 2~m temperature. NeuralGCM has no 2-m diagnostics, and $\Tm$ is its
air temperature at the lowest pressure level, 1000~hPa. Each model is
optimized and evaluated against its own unwarmed control of the same
variable, so absolute temperatures are not compared across models.

\subsection*{Warming, forcing, and target}

The perturbed climate is a uniform $+4$\,K increase of the prescribed
SST. The forcing is a set of five zonal SST bands
(Sec.~\ref{sec:control}), Gaussian in latitude with centers $45\degN$,
$20\degN$, $0^{\circ}$, $20\degS$, $45\degS$ and width
$\sigma_j = 10^{\circ}$, applied over ocean only, each carrying an
amplitude $a_j(t)$ held constant within an optimization segment. The
target is the model's own unwarmed control climatology, so a perfect
controller returns the warmed model to its own unforced climate.

\subsection*{Greedy optimization}

Each segment minimizes the objective of \eqref{eq:loss}, with the
land weights $w$ normalized to sum to one. The amplitude
regularizer weight is $\mu = 0.01$ and the movement penalty is
$\lambda = 0.1$. Each segment is optimized with Adam at learning rate
$0.1$ for 15 iterations, with the gradient averaged over a three-member
ensemble of perturbed initial states, by exact
backpropagation-through-time through the model. The converged
amplitudes are applied, the model integrates through the segment, and
the final state initializes the next segment. No gradient crosses a
segment boundary. The production runs use 91 segments of 8 days
(728 days); the seasonality runs (Section~\ref{si:seasonality}) use
78 segments of 14 days.

\subsection*{Skill and effort}

Runs are scored on four variables -- near-surface temperature,
precipitation, evaporation, and specific humidity, based on their
time-mean fields of the final 364 days of a run.
For variable $v$ the restoration skill is
\begin{equation}
R_v \;=\; 1 \;-\;
  \frac{\bigl\lVert \overline{x}_{\mathrm{opt}} - \overline{x}_{\mathrm{tgt}} \bigr\rVert}
       {\bigl\lVert \overline{x}_{\mathrm{warm}} - \overline{x}_{\mathrm{tgt}} \bigr\rVert} ,
\label{eq:skillR}
\end{equation}
where the overbar is the time mean over the final 364 days, $\lVert \cdot \rVert$ is the
cosine-latitude-weighted spatial RMS over land (grid cells with land
fraction above one half), and the three
trajectories are the optimized run, the target, and the uncontrolled $+4$\,K
run. $R_v = 1$ is exact restoration of the time-mean field and
$R_v = 0$ is no improvement. 

However, restoration is not linear in $R$. The residual error ratio is
$1 - R_v$, so raising $R_v$ from $0.90$ to $0.95$ halves the remaining
error, as large a multiplicative step as raising it from $0.80$ to
$0.90$. The primary skill measure is therefore the gain
\begin{equation}
G_v \;=\; \frac{1}{(1 - R_v)^{2}}
  \;=\; \frac{\bigl\lVert \overline{x}_{\mathrm{warm}} - \overline{x}_{\mathrm{tgt}} \bigr\rVert^{2}}
             {\bigl\lVert \overline{x}_{\mathrm{opt}} - \overline{x}_{\mathrm{tgt}} \bigr\rVert^{2}} ,
\label{eq:gain}
\end{equation}
the factor by which the strategy reduces the mean-square error of the
time-mean field relative to the uncontrolled warmed run. $G_v = 10$
means the error variance is ten times smaller than under no control,
and $G_v = 1$ is no improvement. Equal ratios of $G$ are equal
multiplicative reductions of the residual error, which is the reward
structure of the restoration problem.

The effort is the cooling a strategy prescribes,
\begin{equation}
E \;=\; \Bigl\langle\, \bigl|\, \textstyle\sum_j g_j(\varphi)\, a_j(t) \,\bigr| \Bigr\rangle_{A,\,t} ,
\label{eq:effort}
\end{equation}
the area-weighted global mean and time mean, over the same final
364 days, of the magnitude of the summed prescribed profile, in K, with
$g_j(\varphi)$ being the unit-amplitude Gaussian profiles of Eq.~\eqref{eq:forcingprofile}
before the ocean mask is applied. $E$ measures what the controller asks
for, not what the ocean receives or what a deployment would cost. The
five bands do not tile the sphere, and the area mean of their summed
unit profile is $0.925$, so prescribing a flat $-4$\,K on all five bands
spends $E = 3.70$\,K rather than $4$\,K; only a forcing of $-4$\,K at
every point of the globe would yield $E = 4$.

\subsection*{Cross-model replay}

The strategy learned in JAX-GCM is replayed
into each model, which runs forward without
re-optimization and evaluates the gain (\eqref{eq:gain}) 
over the final 300 days of a one-year replay.
The amplitudes drive each model's own implementation of
the same SST bands of cooling. In LUCIE the SST perturbation is smoothed before
entry (Gaussian smudging, $\sigma = 1.7$ grid cells). Note that the 
near-surface temperature variable differs for different models (Section~\ref{si:transfer}).

\section*{Data, Materials, and Software Availability}
Optimization code, learned forcing strategies, and analysis scripts 
will be deposited in a public repository upon publication.

\section*{Acknowledgments}
PD and AC acknowledge the support from the National Science Foundation 
(grant no. 2425667), the Sloan Foundation, and Schmidt 
Sciences, LLC. DA and AC acknowledges funding from the University of Chicago 
Climate Systems Engineering Initiative (CSEi). Computational resources were 
provided by NSF ACCESS MTH240019, MTH250006 and NCAR CISL UCSC0008, and UCSC0009.

\section*{Author Contributions}
P.K.D., D.S.A., and A.C. designed research; P.K.D. performed research and analyzed data; and P.K.D., D.S.A., and A.C. wrote the paper.

\section*{Competing Interests}
The authors declare no competing interest.

\clearpage
\setcounter{section}{0}
\renewcommand{\thesection}{S\arabic{section}}
\setcounter{figure}{0}
\renewcommand{\thefigure}{S\arabic{figure}}
\setcounter{table}{0}
\renewcommand{\thetable}{S\arabic{table}}
\setcounter{equation}{0}
\renewcommand{\theequation}{S\arabic{equation}}

\begin{center}
{\LARGE\bfseries Supplementary Information}
\end{center}
\bigskip

\section{Gradient validation}\label{si:grad}

A gradient backpropagated through chaotic dynamics grows as $e^{\Lambda T}$
with the horizon $T$, while the true sensitivity of a time-averaged
objective stays bounded, so beyond horizons of order $1/\Lambda$ the
gradient direction decouples from the sensitivity
\citep{lea2000sensitivity, metz2021gradients}. In this section we show that
the backpropagated gradients match those from finite differences
(Fig.~\ref{fig:fdscatter}) for a suitable step $\epsilon$
(Fig.~\ref{fig:fdstep}). There is a weather-dependent uncertainty in the
backpropagated gradients which makes the gradients of longer segments noisy
(Fig.~\ref{fig:icspread}). As a result the noise-to-signal ratio becomes
unfavorable for runs with long segments (Fig.~\ref{fig:seglen}).

\subsection*{Accuracy against finite differences}

\begin{figure*}[t]
\centering
\includegraphics[width=\textwidth]{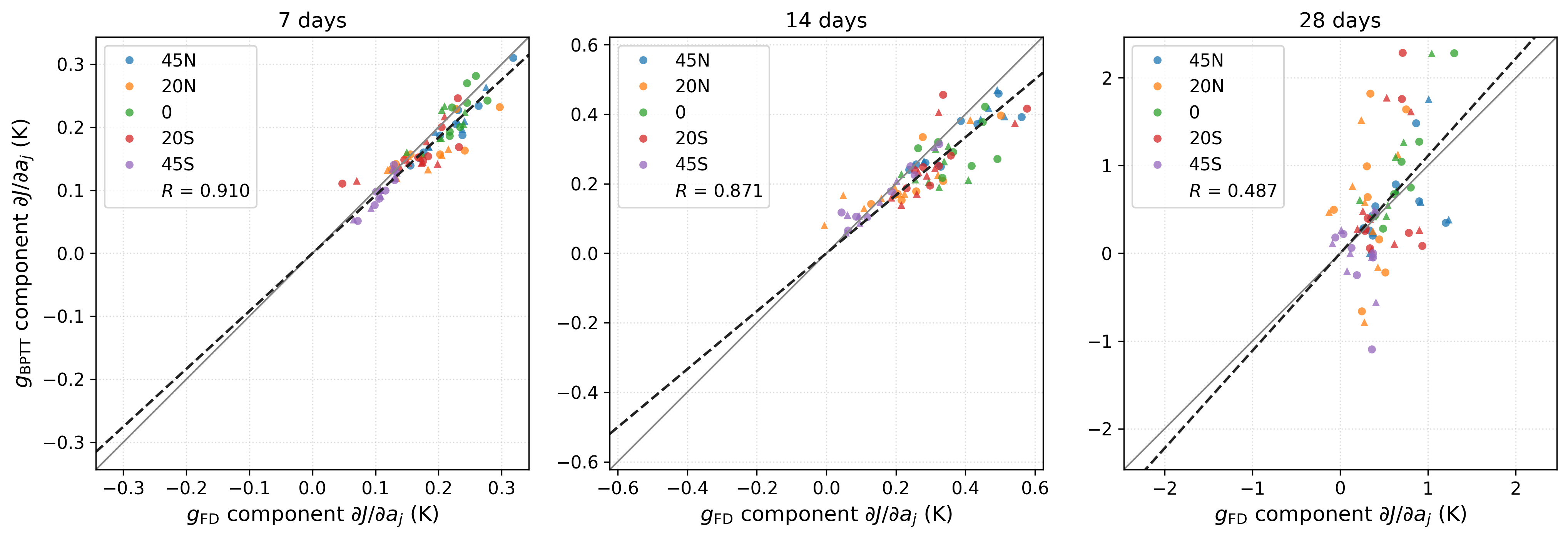}
\caption{Backpropagated gradient ($g_{\mathrm{BPTT}}$) vs. finite-difference gradient
($g_{\mathrm{FD}}$) of the JAX-GCM segment loss. The two agree in
direction and to within $17\%$ in magnitude at 7 and 14 days (correlation
$R = 0.91$ and $0.87$) and decorrelate at 28 days ($R = 0.49$). Each panel
plots the components of the gradient of the loss, $J$, with respect to the
five band amplitudes, one point per band per case and 80 per panel. The
cases are four initial conditions at two operating points, band
amplitudes all zero and all $-2$\,K, under both objectives, with
$\epsilon = 1$\,K on single deterministic trajectories. Colors denote
bands; circles, pattern objective; triangles, land-mean objective. Grey
line, $1$:$1$; dashed line, least-squares fit through the origin; $R$,
correlation of the 80 points. Units, K.}

\label{fig:fdscatter}
\end{figure*}

The segment loss $J(\mathbf{a})$ is the objective $L$ of
Eq.~(\ref{eq:loss}) evaluated for one segment run from a fixed initial
condition with the five band amplitudes $\mathbf{a} = (a_1, \dots, a_5)$
held constant. It is the squared area-weighted land mean of the
segment-mean near-surface temperature difference from the target, plus
$\beta$ times the spatial variance of that difference over land, plus the
amplitude penalty $\mu\,\overline{a_j^2}$, in K$^2$. Throughout this section
$\alpha = 1$ and $\mu = 0.01$, the pattern objective has $\beta = 0.5$ and
the land-mean objective $\beta = 0$, and the movement penalty is absent
($\lambda = 0$) because each case is a single segment with no predecessor.
We compared the backpropagated gradient of $J$ with respect to the five
band amplitudes against central finite differences on single deterministic
trajectories, at segment lengths of 7, 14, and 28 days, from four
initial conditions (January 1, April 1, July 1, and October~1), at two operating
points (band amplitudes all zero and all $-2$\,K), and under both
objectives, for 16 cases per segment length. For each case we perturbed one
band amplitude at a time by $\pm 1$\,K about the operating point, held the
other four fixed, re-ran the segment from the same initial condition, and
formed the five-component central-difference gradient
\[
  \left(g_{\mathrm{FD}}\right)_j
  = \frac{J(\mathbf{a} + \epsilon \mathbf{e}_j)
          - J(\mathbf{a} - \epsilon \mathbf{e}_j)}{2\epsilon},
  \qquad \epsilon = 1\ \mathrm{K},
\]
where $\mathbf{e}_j$ is the unit vector of band $j$.
Fig.~\ref{fig:fdscatter} plots all five components of all 16 cases against
the backpropagated gradient, $g_{\mathrm{BPTT}}$, at each segment length. At
7 and 14 days the components lie along the $1$:$1$ line with correlations
of $0.91$ and $0.87$, and a least-squares fit through the origin has slopes
of $0.92$ and $0.83$, so the backpropagated gradient points the right way with a magnitude
 $8$ to $17\%$ smaller than finite difference. The median angle between the two gradient
vectors is $5.2^{\circ}$ and $7.6^{\circ}$ under the pattern objective and
$5.7^{\circ}$ and $9.0^{\circ}$ under the land mean objective. 
At 28 days the points spread out and correlation falls to $0.49$.
The median angle between FD and BPTT gradient vectors 
rises to $45^{\circ}$ under the
pattern objective and $41^{\circ}$ under the land mean, the cases range
from $9^{\circ}$ to $86^{\circ}$, and  sign disagrees at least sometmes in three
components out of five. 

\subsection*{Choice of the finite-difference step}

\begin{figure*}[t!]
\centering
\includegraphics[width=\textwidth]{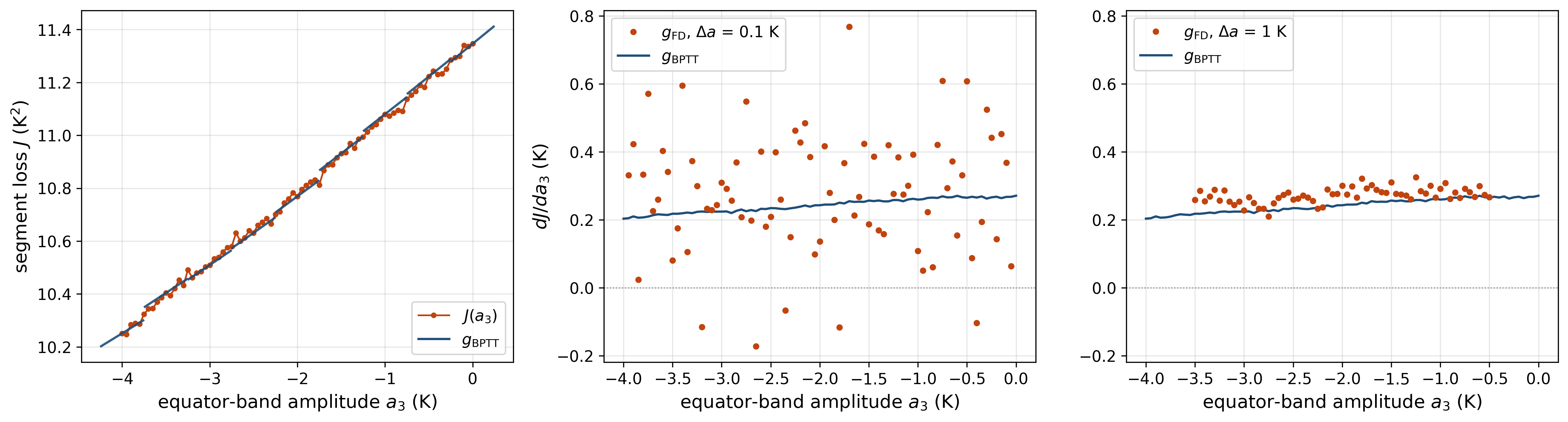}
\caption{Finite-difference slope of the segment loss against the
differencing step in JAX-GCM. Left: segment loss, $J$, of the land-masked
pattern objective as a function of the equator-band amplitude $a_3$. Here we plot 81 evaluations
between $0$ to $-4$\,K in $0.05$\,K steps for an 8-day segment with the other four amplitudes
set to zero. We also plot the backpropagated tangent for at every tenth point as a line. 
Middle and right: points denote finite-difference slopes
$g_{\mathrm{FD}} = [J(m + \Delta a/2) - J(m - \Delta a/2)]/\Delta a$ over
pairs of sampled amplitudes $\Delta a = 0.1$\,K and $1$\,K apart, plotted at
their midpoint $m$; line denotes the backpropagated $\dd J/\dd a_3$.}
\label{fig:fdstep}
\end{figure*}

A central difference of a smooth function converges as $\epsilon \to 0$,
but here the step is bounded from below by weather noise.
Fig.~\ref{fig:fdstep} (left) shows the segment loss, $J$, along the equatorial band
amplitude, $a_3$, at $0.05$\,K resolution for an 8-day segment under the
land-masked pattern objective, with the other four amplitudes set to zero. 
The loss
follows a  nearly linear trend that the backpropagated tangent
tracks, with a scatter of $\sigma_J = 0.014$\,K$^2$ about it ($1.2\%$ of
the range of $J$) because each evaluation runs a different trajectory. A
finite difference over a step $\Delta a$ inherits that scatter as
$\sqrt{2}\,\sigma_J / \Delta a$. At $\Delta a = 0.1$\,K the
finite-difference slopes scatter by $\pm 0.18$\,K about a mean of
$0.28$\,K such 5 of 79 have the opposite sign (negative) (Fig.~\ref{fig:fdstep}, middle); at
$\Delta a = 1$\,K they scatter by $\pm 0.02$\,K and converge onto a much smoother
curve $12\%$ above the backpropagated gradient (Fig.~\ref{fig:fdstep}, right),
the same shortfall as the fit slopes of Fig.~\ref{fig:fdscatter} at 7 and
14 days. The $1$\,K step of Fig.~\ref{fig:fdscatter} is set by this noise
floor and not by the behavior of the gradient.

\subsection*{Dependence on the weather realization}

\begin{figure*}[t!]
\centering
\includegraphics[width=\textwidth]{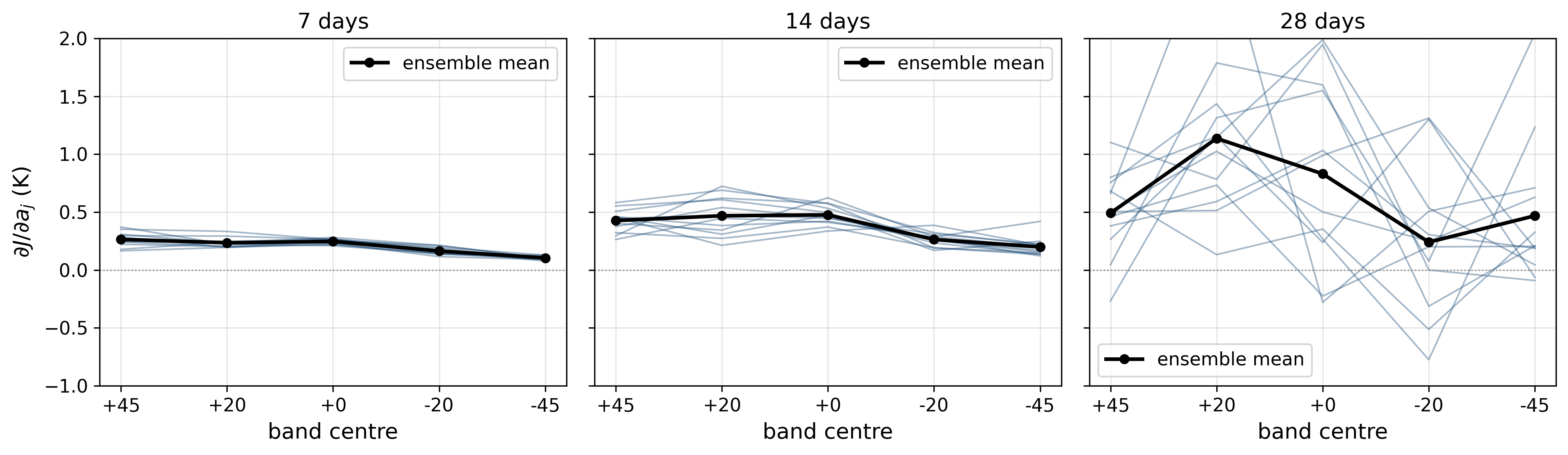}
\caption{Weather dependence of the backpropagated segment gradient in
JAX-GCM. The gradient of a single weather realization stays within
$9^{\circ}$ of the twelve-realization mean up to 14 days and departs by
$41^{\circ}$ at 28 days. Backpropagated gradient of the pattern objective
with respect to the five band amplitudes, at all amplitudes set to zero, from
twelve weather realizations (faint lines) and their mean (black), at
segment lengths of 7, 14, and 28 days from left to right on a common
vertical axis. Initial conditions are sampled every 7 days over a 77-day
span of the $+4$\,K warmed band-free run. Units, K.}
\label{fig:icspread}
\end{figure*}

Fig.~\ref{fig:fdscatter} compares backpropagated and finite-difference
gradients of the same trajectory, so the weather dependence of the gradient
is not relevant. 
To measure the effect of weather we compare the
backpropagated gradient of the pattern objective, with all band amplitudes
at zero, across twelve initial conditions drawn from the $+4$\,K warmed
band-free run (Fig.~\ref{fig:icspread}). The first is the spun-up
January~1 state that also starts the January cases of
Fig.~\ref{fig:fdscatter}. Each subsequent one is the previous state advanced
7 days with the band amplitudes held at zero, so the twelve are consecutive
states of one trajectory, 7 days apart and 77 days end to end.
Each faint line is one backpropagated gradient vector
$g_{\mathrm{BPTT}} = (\partial J/\partial a_1, \dots, \partial J/\partial a_5)$
from one weather realization, drawn against the band centre, and the thick
black line is the mean of the twelve.

At 7 and 14 days every
realization returns a positive gradient on all five bands, and the standard
deviation across realizations is $9$ to $23\%$ of the mean component at 7
days and $18$ to $40\%$ at 14 days. As an angle, a single realization's
gradient sits $5.5^{\circ}$ from the twelve-member mean at 7 days and
$9.0^{\circ}$ at 14 days, after removal of a linear drift over the 77-day
span, and $41^{\circ}$ at 28 days, where 8 of the 12 realizations return a
negative component on at least one band. A descent step along a unit vector
at angle $\theta$ to the true gradient reduces the loss by $\cos\theta$ of
the achievable amount, so $9.0^{\circ}$ at 14 days costs $1.2\%$ of the
available reduction per segment and $41^{\circ}$ at 28 days costs $24\%$.


\subsection*{Noise-to-signal ratio in the production runs}

\begin{figure*}[t!]
\centering
\includegraphics[width=\textwidth]{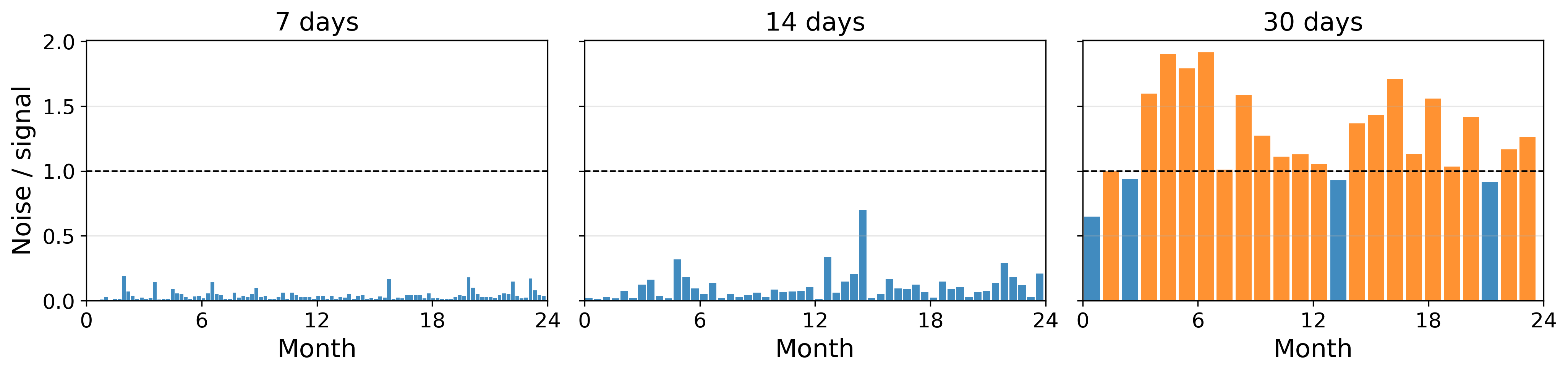}
\caption{Noise-to-signal ratio of the production gradient against segment
length in JAX-GCM. Every segment stays below one at 7 and 14 days and most
exceed one at 30 days. Per-segment ratio of the norm of the cross-member
standard deviation of the gradient to the norm of the ensemble-mean
gradient, averaged over the segment's iterations, for three runs of the
five-band zonal control problem over two years at segment lengths of 7, 14,
and 30 days (104, 52, and 24 segments; land-masked pattern objective with
$\alpha = 1$ and $\beta = 0.5$, $0.05$, and $0.1$; amplitude regularizer
weight 0.01; three-member ensembles; 15 Adam iterations per segment). Blue,
below one; orange, above; dashed line, noise equal to signal.}
\label{fig:seglen}
\end{figure*}

For segment lengths of 7, 14, and 30 days over two years (104, 52, and 24
segments) the optimizer evaluates the gradient of each segment from three
initial conditions separated by a small perturbation. Each member returns
its own gradient of the segment loss, and the run logs the norm of the
ensemble-mean gradient (the signal) and the norm of the component-wise
standard deviation across members (the noise). Fig.~\ref{fig:seglen}
plots the ratio of the two averaged over the 15 iterations of each segment.
At 7 and 14 days every segment stays below one (medians $0.03$ and $0.07$,
maxima $0.19$ and $0.70$). At 30 days the median is $1.2$ and noise exceeds
signal in 20 of 24 segments. The crossing between 14 and 30 days matches the
breakdown between 14 and 28 days in Figs.~\ref{fig:fdscatter} and
\ref{fig:icspread}.

The cooling degrades later than the gradient. Over the final 364 days of
each run the optimizer removes $89\%$ of the land-mean warming at 7 days,
$95\%$ at 14 days, and $95\%$ at 30 days (the 7-day run has $\beta = 0.5$
against $0.05$ and $0.1$ for the other two, so its fraction is not directly
comparable), and a 50-day run not shown in Fig.~\ref{fig:seglen}, with
every segment above one and a median ratio of $1.8$, removes $46\%$. The
30-day run therefore reaches its target while the noise exceeds the signal
in 20 of its 24 segments. 

\section{Validation in LUCIE and NeuralGCM}\label{si:emuval}

The optimization of Sec.~\ref{sec:control} runs unchanged in LUCIE and
NeuralGCM. Each model optimizes its own five band amplitudes through
its own dynamics, toward its own unwarmed climate as target, with the
land-masked objective of JAX-GCM (the pattern loss with
$\beta = 0.5$, or the land-mean loss), 7- or 8-day segments,
and the same amplitude regularizer. The runs shown are single
realizations with zero movement penalty.

\begin{figure*}[t!]
\centering
\includegraphics[width=\textwidth]{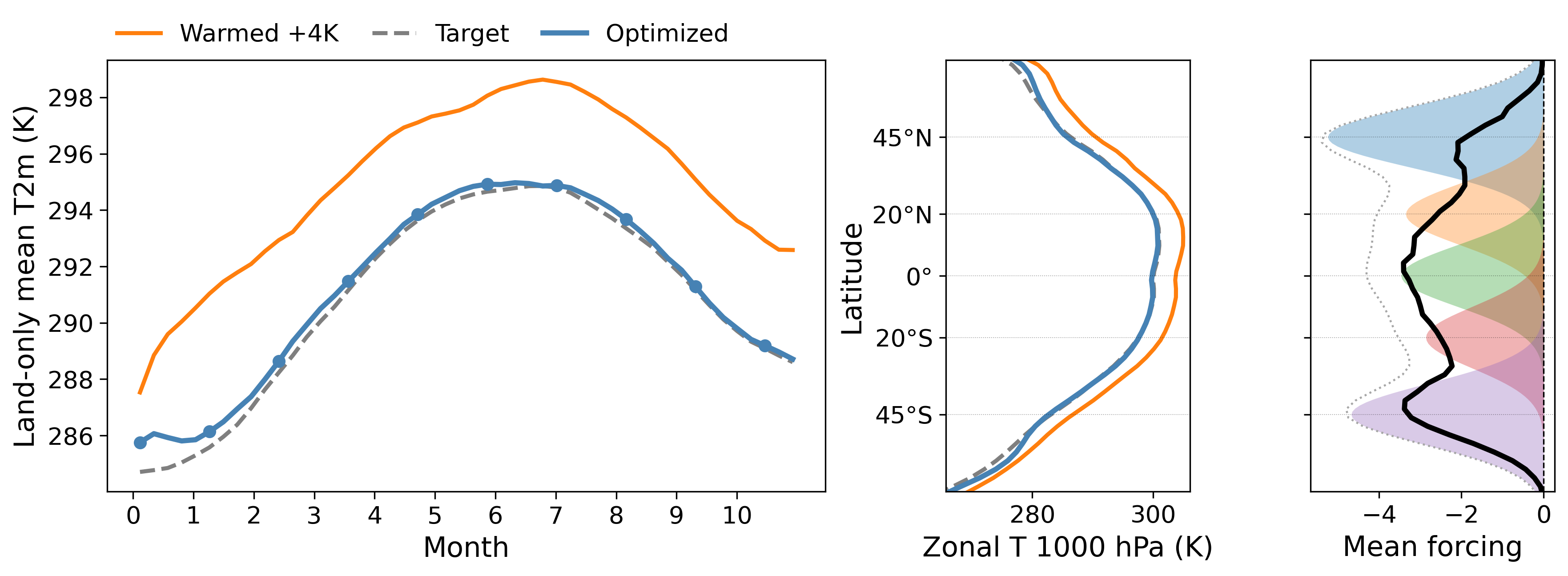}
\caption{NeuralGCM optimizing its own band amplitudes under the pattern
loss over one year (52 segments of 7 days, three-member gradient
ensemble, no movement penalty). Left: land-mean $\Tm$ (K) against
time (months) for the warmed ($+4$\,K), target, and optimized runs;
dots mark every fifth segment. Middle: zonal-mean temperature at
1000\,hPa (K) against latitude for the same three runs, time means over
the final 364 days. Right: time-mean applied cooling (K) against
latitude for the ocean-fraction- and area-weighted forcing
profile of \eqref{eq:forcingprofile} (thick black), the raw amplitude
profile (thin dotted), and the five Gaussian bands (filled curves).}
\label{fig:ngcm-control}
\end{figure*}

\begin{figure*}[t!]
\centering
\includegraphics[width=\textwidth]{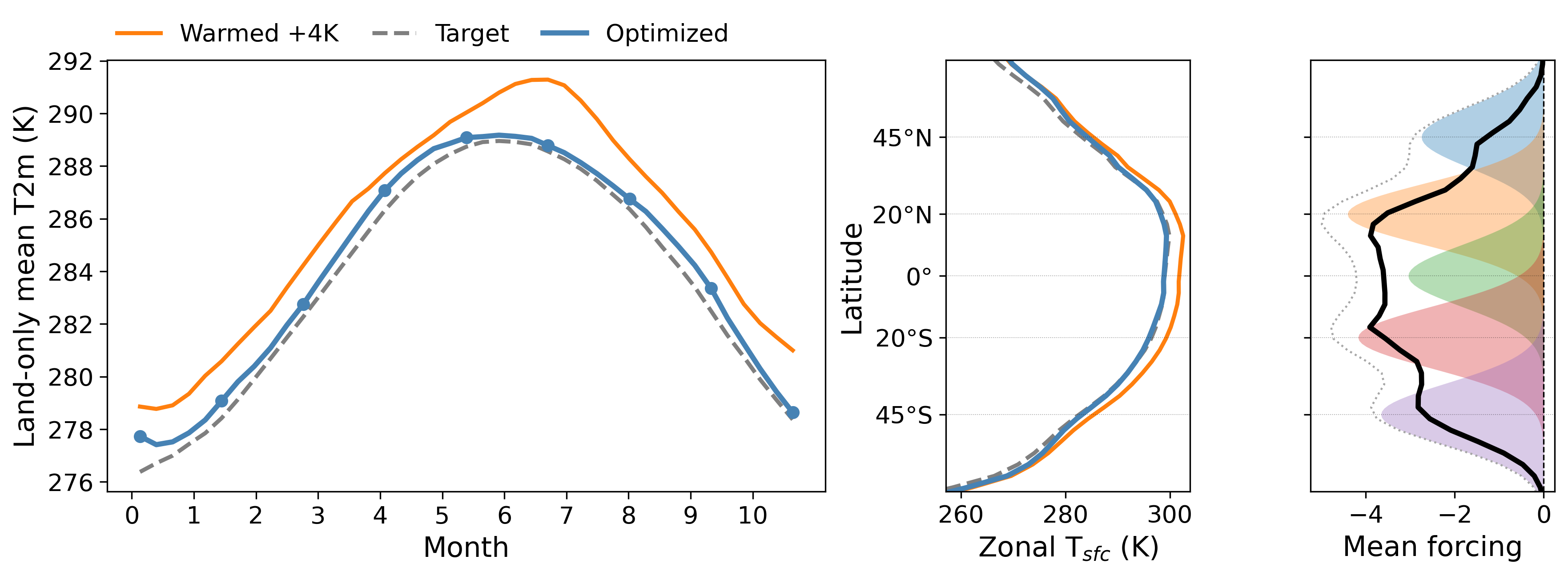}
\caption{LUCIE optimizing its own band amplitudes under the pattern loss
over one year (45 segments of 8 days, three-member gradient ensemble,
no movement penalty). Panels as in Fig.~\ref{fig:ngcm-control}, with
the middle panel showing zonal-mean surface temperature (K).}
\label{fig:lucie-control}
\end{figure*}

Our method controls both models. Under the pattern objective each
model's optimized run converges within a few segments and then
tracks its own target seasonal cycle
(Figs.~\ref{fig:ngcm-control} and~\ref{fig:lucie-control}), removing
92\% (NeuralGCM) and 82\% (LUCIE) of the land-mean warming realized in
the corresponding $+4$\,K run over one year. The land-mean runs remove
90\% and 91\% of the land-mean warming (not shown).

Under the pattern objective NeuralGCM reproduces the JAX-GCM strategy,
placing its largest time-mean cooling on the $45\degN$ band with a
secondary maximum at $45\degS$ and weaker forcing at lower latitudes 
(Fig.~\ref{fig:forcing-pattern}).
In contrast, LUCIE puts its largest time-mean amplitudes
on the subtropical bands, $-4.4$\,K at $20\degN$ and $-4.2$\,K at
$20\degS$, and its $45\degN$ amplitude is the shallowest of its five,
$-2.7$\,K (Fig.~\ref{fig:forcing-pattern}). 
LUCIE's disagreement may be the result of poor behavior for out-of-distribution
SST forcing \citep{merchant2026zones}.

\section{Cross-model transfer of the learned strategy}\label{si:transfer}

\begin{figure*}[t!]
\centering
\includegraphics[width=0.94\textwidth]{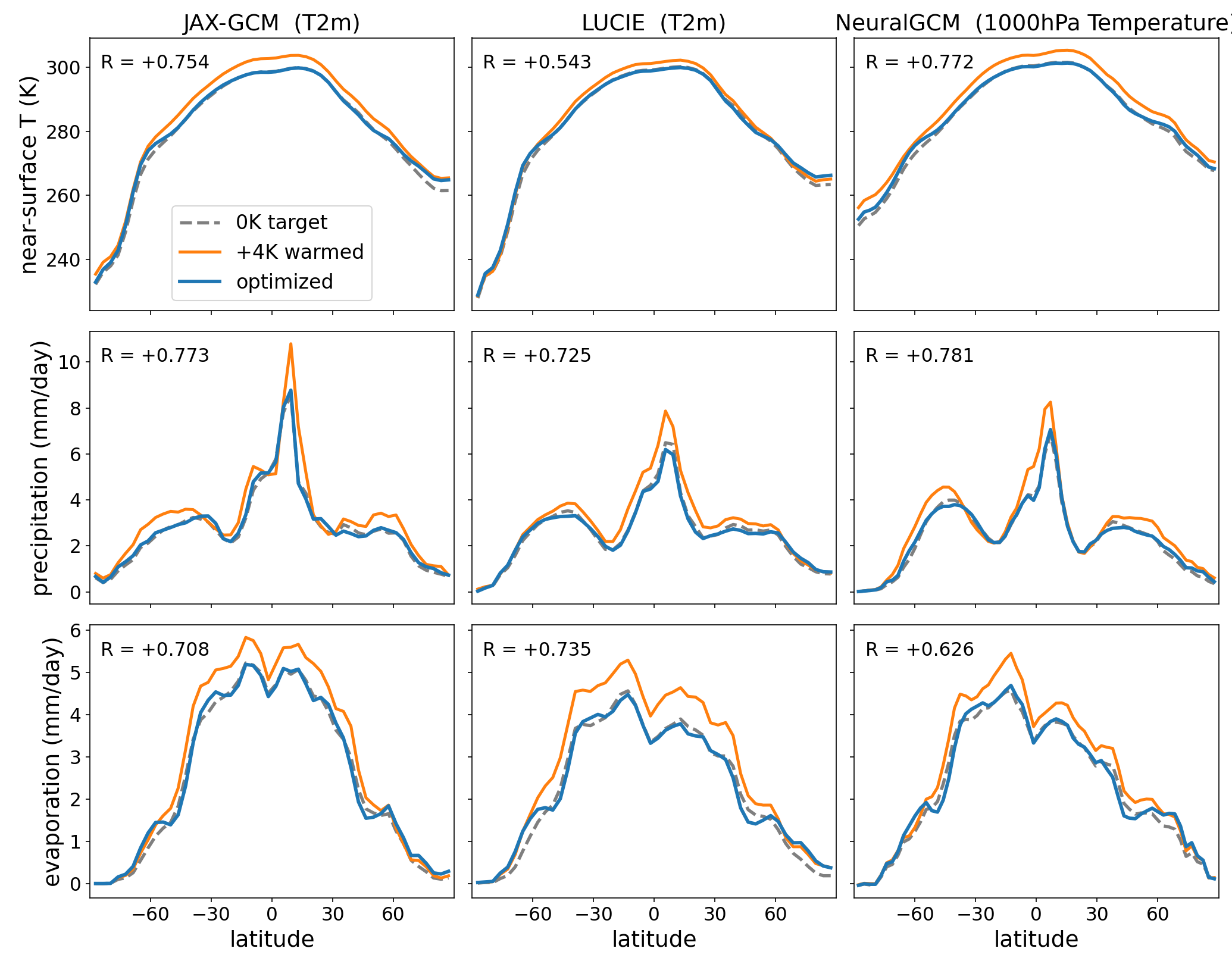}
\caption{Zonal-mean, time-mean profiles over days 60--360 of the
replay year, one column per host, for near-surface temperature (top,
K), precipitation (middle, mm\,day$^{-1}$), and evaporation (bottom,
mm\,day$^{-1}$). Lines represent the host model's own unwarmed control (grey dashed),
the uncontrolled $+4$\,K run (orange) and the frozen JAX-GCM strategy replayed
in the host (blue). The annotated $R$ is the zonal-profile restoration skill
defined in the text.}
\label{fig:replay-zonal}
\end{figure*}

Figure~\ref{fig:replay-zonal} shows the zonal-mean structure behind the
bars of Fig.~\ref{fig:replay} computed from the same replay. 
The 8-day JAX-GCM strategy of Fig.~\ref{fig:longrun} is run forward in JAX-GCM, 
LUCIE, and NeuralGCM with no re-optimization, and each column compares that
model's controlled run against its own unwarmed control and its own
uncontrolled $+4$\,K run. The profiles are time-averaged over 300 days, with 
the first 60 days discarded as transient.

Near-surface temperature is a different quantity in each model. JAX-GCM
has no 2-m diagnostic and interpolates it from the skin temperature and
SPEEDY's surface air temperature $t_0$ as
$\Tm = T_{\mathrm{skin}} + 0.585\,(t_0 - T_{\mathrm{skin}})$, LUCIE
carries the ERA5 2-m temperature as a model variable, NeuralGCM reports
its lowest-level air temperature at 1000~hPa. Absolute temperatures should
therefore not be compared across columns, and when NeuralGCM optimizes for
itself (Section~\ref{si:emuval}) its objective is its 1000-hPa
temperature rather than the JAX-GCM $\Tm$. Precipitation and evaporation
are in mm\,day$^{-1}$ in all models.

The annotated $R$ is the restoration skill of the time-mean zonal
profile, $R = 1 - \mathrm{RMS}(\mathrm{controlled} - \mathrm{control}) /
\mathrm{RMS}(\mathrm{warmed} - \mathrm{control})$, with a
cosine-latitude-weighted RMS over latitude, the zonal-profile
counterpart of the field skill $R_v$ of the Materials and
Methods. $R = 1$ is full restoration and $R = 0$ is no restoration with respect
to the warmed initial state.

Only $\Tm$ over land entered the loss that produced the strategy, and
only in JAX-GCM. In all three models the replayed strategy restores the
zonal temperature profile as well as the
 the precipitation and evaporation profiles, which were not included in the loss function. 
Restoration across models including variables not included in the loss function 
indicates that the strategy corrects the forced response and not the 
temperature field alone.

\section{Optimizer variants and baseline construction}\label{si:variants}

\begin{figure*}[t!]
\centering
\includegraphics[width=\textwidth]{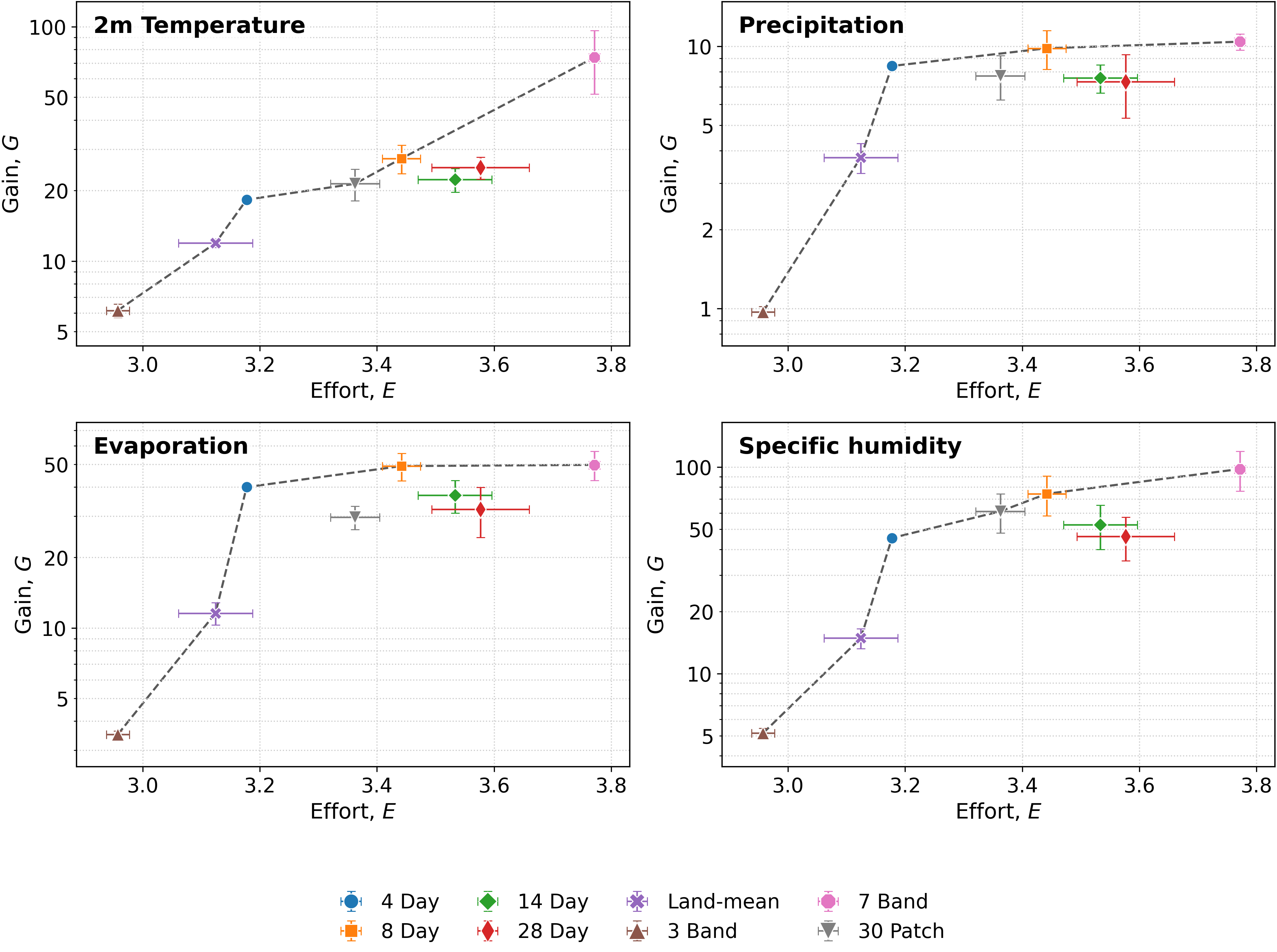}
\caption{Restoration gain $G$ against prescribed effort $E$ (K) for
variants of the optimizer in JAX-GCM, one panel per
variable, computed over the final 364 days of each 728-day run. The
variants are the segment lengths (4, 8, 14, and 28 days) under the pattern
objective, the land-mean objective, three- and seven-band
actuators, and an
actuator of 30 two-dimensional ocean Gaussian patches, the five bands
split into six longitudes each. Markers are
ensemble means where more than one member exists (8- and 14-day ten
members, 28-day four, Land-mean and 3 Band five, 7 Band and Patch
three) with
bars one standard deviation; the 4 Day variant is a single run. The dashed
line joins the strategies not dominated in that panel.}
\label{fig:pareto-variants}
\end{figure*}

This section constructs the baselines of Fig.~\ref{fig:pareto}
(Sec.~\ref{sec:pareto}) and compares variants of the optimizer on the
same effort--gain plane (Fig.~\ref{fig:pareto-variants}). 
The effort $E$ is the 
prescribed cooling $\sum_j g_j(\varphi)\,a_j(t)$ over the 
final 364 days with $g_j$ being the Gaussian
band profiles and $a_j(t)$ the amplitudes
(Eq.~\eqref{eq:effort}). The gain $G$ is the measure
of Fig.~\ref{fig:pareto} (\eqref{eq:gain}) evaluated per variable over the final 364 days.
 $R_v$ is the fractional reduction of the cosine-latitude-weighted
spatial RMS error over land of the time-mean field relative to the uncontrolled
$+4$\,K run, $G_v = 1/(1-R_v)^2$ is the factor by which an intervention reduces
the mean-square error of the time-mean field
(Eqs.~\eqref{eq:skillR} and~\eqref{eq:gain}).

The 8-Day pattern strategy (ten-member ensemble, pattern-loss configuration, Sec.~\ref{sec:control})
and the land-mean strategy (five-member ensemble, land-mean-objective Sec.~\ref{sec:lossdesign}) are learnt. 
The $-4$\,K baseline applies $-4$\,K on all five bands. 
The matched-effort baseline has an amplitude of $-3.72$\,K for all bands,
chosen so that its effort equals the learned ensemble's ($E = 3.44$\,K in
both). Its amplitude is set by the learned result, so it is not an
independent design; at fixed effort it isolates what the spatial
structure of the forcing contributes. The matched-pattern baseline
holds the learned pattern strategy's time-mean amplitudes constant and
isolates what the time variation contributes.
Each flat baseline is a five-member ensemble generated by varying the
spin-up length from one to five years under identical forcing.

The linear baseline replaces the search over amplitudes with a single
inversion of a response operator, and stands in for the optimal-pattern
and response-function methods of Section~\ref{si:prior}
\citep{banweiss2010optimization, lu2020neutral}. Collect the residual field
$d$ of \eqref{eq:loss} at the $n = 4608$ grid points in $\mathbf{d}$, and the land
weights, normalized to sum to one, in $\mathbf{w}$. The two temperature terms
of \eqref{eq:loss} are then a single quadratic form $\mathbf{d}^{\!\top} M
\mathbf{d}$, with
\begin{equation}
  M \;=\; \beta\,\mathrm{diag}(\mathbf{w})
     \;+\; (\alpha - \beta)\,\mathbf{w}\mathbf{w}^{\!\top},
\label{eq:si-lin-M}
\end{equation}
the rank-one term carrying the squared land-mean bias and the diagonal term
the spatial variance of the residual. Modeling the residual as linear in the
amplitudes about the uncontrolled $+4$\,K state, $\mathbf{d}(\mathbf{a})
\approx \mathbf{d}_0 + \mathcal{G}\,\mathbf{a}$, where $\mathbf{d}_0$ is the
residual of that uncontrolled run and column $j$ of the $n \times k$ operator
$\mathcal{G}$ is the time-mean $\Tm$ response per unit amplitude on band $j$,
the minimizer of \eqref{eq:loss} follows in closed form,
\begin{equation}
  \mathbf{a}^{\star} \;=\; -\Bigl(
    \mathcal{G}^{\!\top} M\, \mathcal{G} + \frac{\mu}{k}\,I
  \Bigr)^{-1} \mathcal{G}^{\!\top} M\, \mathbf{d}_0 ,
\label{eq:si-lin-solve}
\end{equation}
where $\mathcal{G}$ is the response operator and not the gain $G$ of
\eqref{eq:gain}. The matrix inverted in \eqref{eq:si-lin-solve} is $k \times k$
with $k = 5$ bands. The amplitude regularizer of \eqref{eq:loss} is a mean over the
five bands, so it enters \eqref{eq:si-lin-solve} as $\mu/k$, and the
movement penalty $\lambda$ vanishes for amplitudes held constant.

Each column of $\mathcal{G}$ was estimated from forward runs by central
differences, $\mathcal{G}_j = [\mathbf{d}(+\varepsilon\mathbf{e}_j) -
\mathbf{d}(-\varepsilon\mathbf{e}_j)]/2\varepsilon$ with $\varepsilon =
4$\,K, following the Green's-function construction
\citep{hassanzadeh2016lrf}. Each entry is a difference of two 364-day time
means, and each perturbation run is a 24-member ensemble over perturbed initial states in which member $i$ of
the $+\varepsilon$ run shares its initial state with member $i$ of the
$-\varepsilon$ run. The operator took 288 forward integrations of 728 days
and no gradient. The full nonlinear model was then run at the fixed
$\mathbf{a}^{\star}$, as a five-member spin-up ensemble like the flat
baselines. The response is not linear across $[-\varepsilon, +\varepsilon]$.
Warming the tropical ocean produces about twice the land-weighted mean
response per K that cooling it produces, so $\mathbf{a}^{\star}$ depends on
which secant estimates $\mathcal{G}$. This baseline is a linear-response
reference rather than a converged optimum. A comparison of the gain obtained
via the optimizer against the baselines discussed above is shown in 
Fig.~\ref{fig:pareto}.

Fig.~\ref{fig:pareto-variants} varies the optimizer rather than the
baseline, with the segment length at 4, 8, 14, and 28 days, the
land-mean objective, a three-band actuator (centers $45\degN$,
$0^{\circ}$, $45\degS$), a seven-band actuator with centers every
$20^{\circ}$ from $60\degS$ to $60\degN$, and a patch actuator that
opens the five bands to longitudinal variation, six patches 
Among the segment
lengths, overall skill (the four-variable mean of $R_v$) is highest
for the 8-day variant. On 2\,m temperature the 8-, 14-, and 28-day
segments do nearly equally well, but
the longer segments fall off on the other variables and the
8-day segment leads consistently across all variables.
Raising the amplitude regularizer ($\mu$) far above its adopted weight 
shrinks the prescribed amplitudes and the associated restoration skill
toward zero. At $\mu = 0.5$, fifty times the adopted weight (with
14-day segments), the prescribed effort falls to $E = 1.1$\,K and the gain to $G = 2.2$ on
2\,m temperature and $G = 2.2$ on precipitation, averaged over three
ensemble members (not shown in Fig.~\ref{fig:pareto-variants}).

Since the gradients of the loss function with respect to all control
parameters are obtainable in a single backward pass, the optimization 
algorithm scales well to high-dimensional control and the computational 
cost of the 30-patch problem remains similar to the 5-band problem.
However, the added degrees of freedom does not aid the optimization as
the patch basis shows less skill than the 5-band basis across all
variables. This may be due to the design of the objective function 
or convergence toward a local solution.


While the increased degrees of freedom in the patch basis do not improve
the optimizer skill, increasing the number of zonal bands has an outsized
effect. For the 7-band basis, although the precipitation and evaporation
gains saturate, the 2\,m temperature gain increases by a factor of 2.7 and
the specific humidity gain by 1.3.
Increasing the effort in the 5-band basis (via decreasing $\mu$ in the 
loss function) to match that of the 7-band solution yields only marginal
improvements in the restoration skill (not shown). Thus the increased band count and
greater coverage over the ocean area are the primary drivers for high gain in 
the 7-band basis. Conversely, replacing the five bands with three has a 
large adverse effect as the precipitation gain falls to 1 (no improvement)
and the evaporation and humidity gains to 3.5 and 5, while still removing
about three quarters of the land-mean warming. This suggests that latitudinal structure as detailed as the physical cooling mechanism (marine cloud brightening or stratospheric aerosols) allows should be exploited.

Sweeps not shown in Fig.~\ref{fig:pareto-variants} leave the result
unchanged. Across the optimizer (Adam against L-BFGS-B, and
best-iterate against last-iterate selection), the movement penalty in
weight and norm, and the amplitude regularizer within its adopted
range, the five-band amplitude structure keeps a similar shape, and the ensemble-mean restoration skill is unchanged to
within the member-to-member scatter. The robustness statement applies 
to the neighborhood of the adopted operating point, not to arbitrary 
parameter values.

\begin{figure*}[t!]
\centering
\includegraphics[width=\textwidth]{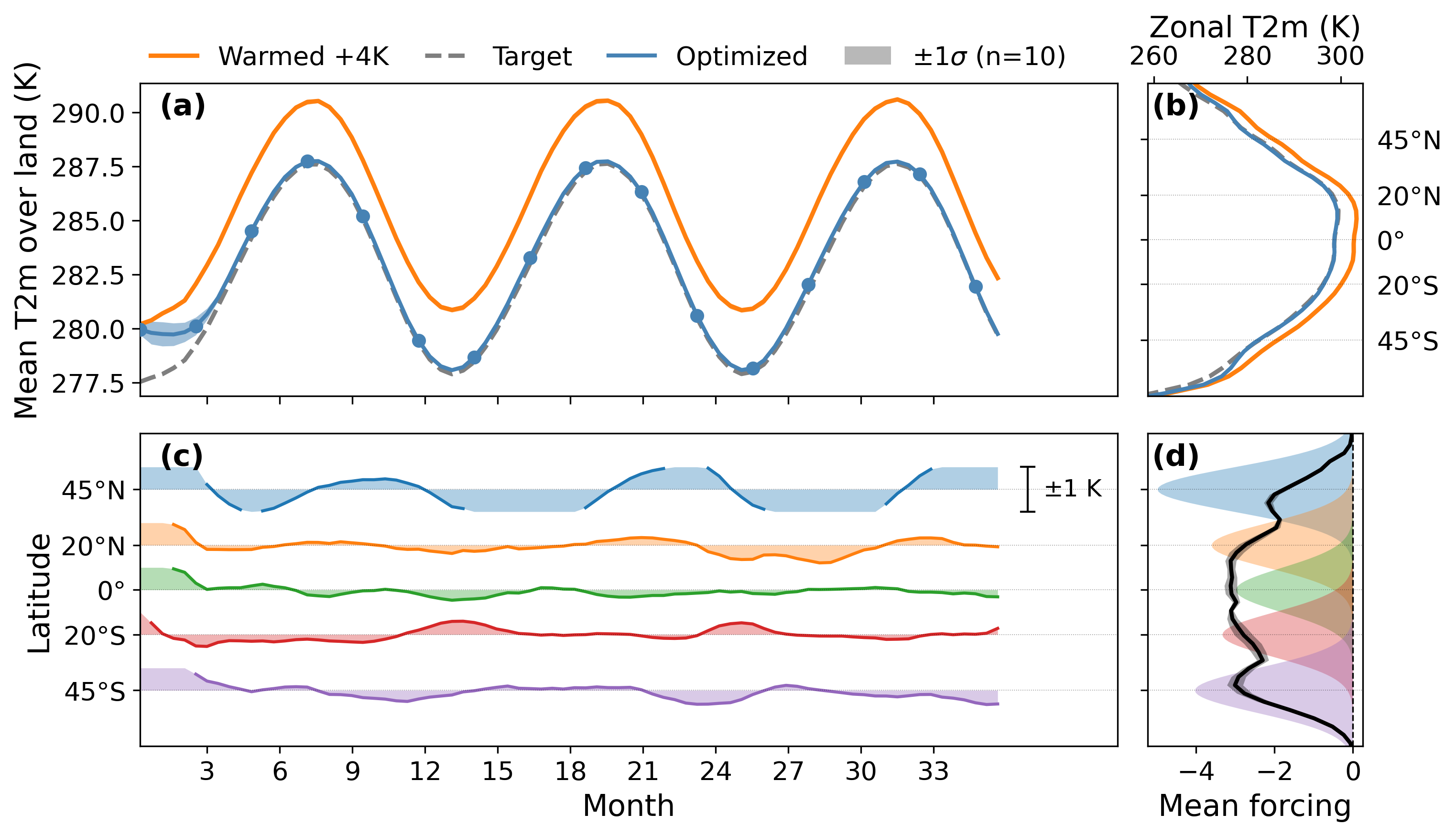}
\caption{Three-year control in JAX-GCM at the 14-day
optimization horizon, a ten-member ensemble with the loss and
penalty of the Fig.~\ref{fig:longrun} ensemble, 78 segments of 14 days. All time series
plots are shown as a moving average over 5 segments (70 days).
(a)~Land-mean $\Tm$ (K) against time (months) for the warmed
($+4$\,K), target, and optimized runs. The optimized curve shows every
fifth segment with a dot. (b)~Zonal-mean $\Tm$ (K) against latitude for
the same three runs. (c)~Temporal variations in the learned strategy.
Each band's cooling amplitude is shown as the deviation from its
time-averaged value in (d), drawn about the band's own latitude; the bar
at the upper right gives the scale, $\pm 1$\,K. (d)~Time-averaged values
of the applied cooling against latitude, with the average taken over the
full three-year run. The filled curves denote the five Gaussian bands
with the amplitudes indicated by their heights. Thick black line shows
the ocean-fraction and area-weighted forcing profile of Eq.~S3.}
\label{fig:control14}
\end{figure*}

\section{Learned amplitudes and seasonality}\label{si:seasonality}

\begin{figure*}[t!]
\centering
\begin{subfigure}[t]{0.48\textwidth}
\centering
\includegraphics[width=\textwidth]{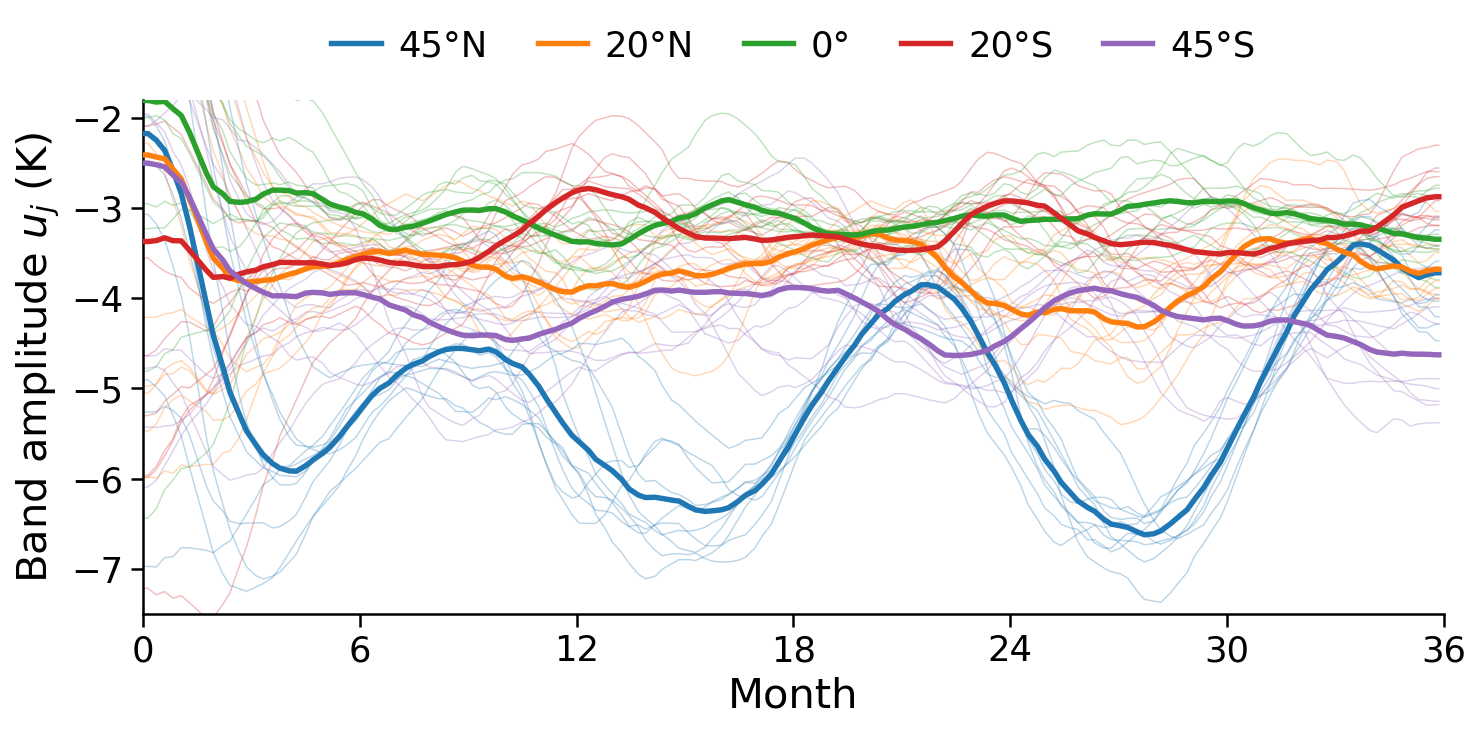}
\caption{}
\label{fig:seasonality-native}
\end{subfigure}%
\hfill
\begin{subfigure}[t]{0.48\textwidth}
\centering
\includegraphics[width=\textwidth]{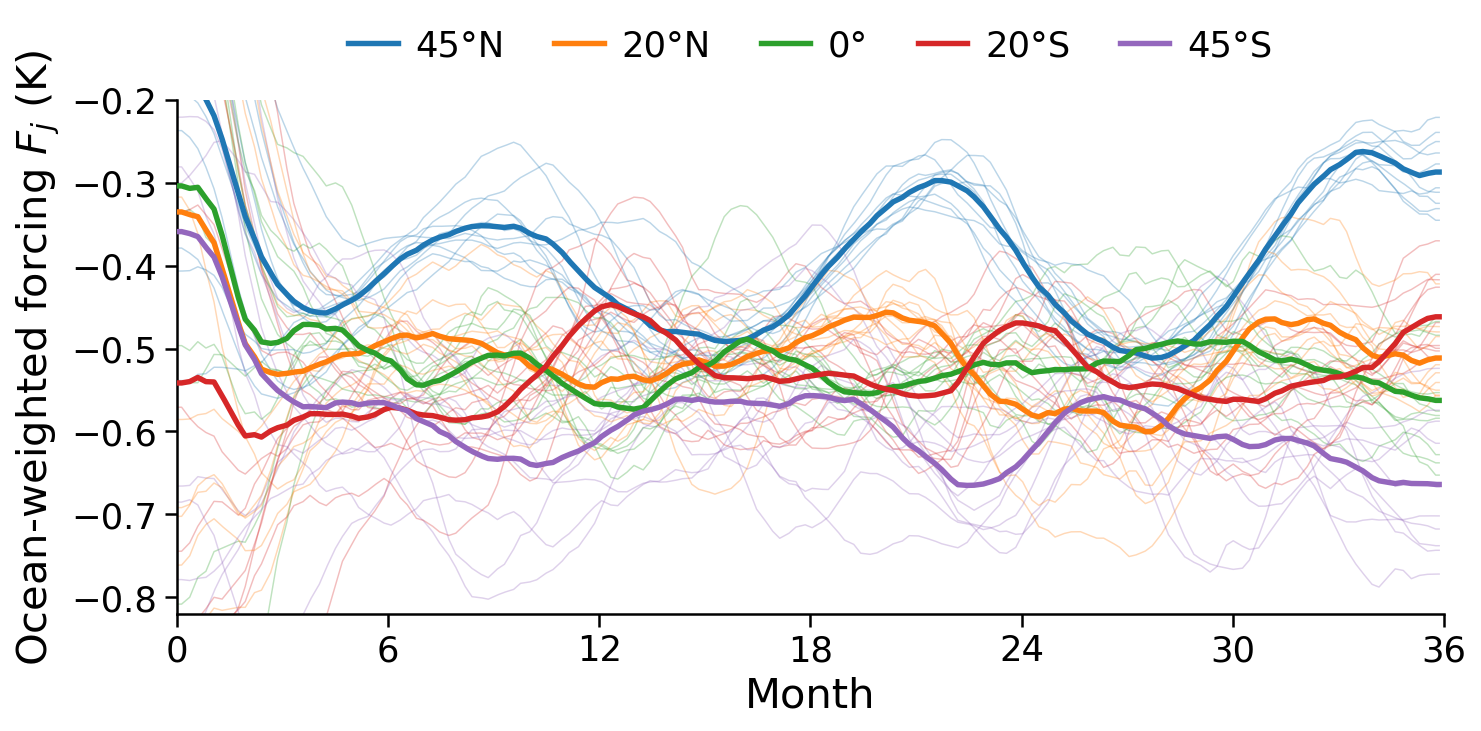}
\caption{}
\label{fig:seasonality-forcing}
\end{subfigure}
\caption{The learned strategy over three years in JAX-GCM for a ten-member ensemble
run with identical settings. The ten thin lines are the individual
members, the thick line is the ensemble mean,  and both are drawn under a 90-day centered moving average.
(a)~Prescribed amplitude (K) of each of the five Gaussian cooling bands
against time (months). (b)~The ocean-area-weighted forcing
$F_j(t) = a_j(t)\,A_j$ each band delivers to the global-area mean (K,
\eqref{eq:bandforcing}), where $A_j$ is band $j$'s effective ocean area as a fraction of
global area. Note the change of vertical scale between the panels.}
\label{fig:seasonality}
\end{figure*}

The 14-day ensemble of Fig.~\ref{fig:control14} has ten members initialized with
different band amplitudes, with 
78 segments of 14 days, the pattern loss, and
the movement penalty $\lambda = 0.1$. 
Fig.~\ref{fig:seasonality} shows the learned
band amplitudes and forcings of this ensemble.

The optimizer does not prescribe equal amplitudes across bands
(Fig.~\ref{fig:seasonality-native}) as the $45\degN$ band maintains an
amplitude about 1-2\,K larger than every other band. The realized applied
forcing $F_j(t) = a_j(t) A_j$ of
\eqref{eq:bandforcing} (Fig.~\ref{fig:seasonality-forcing}) has a lower
spread between time-averaged amplitude means.  The $45\degN$ band with the 
largest amplitude has the lowest applied forcing and the $45\degS$ band
with greatest ocean fraction has the highest applied forcing.

The $45\degN$ and the $45\degS$ bands carry a clear seasonal cycle and 
are anti-phase with each other, but for the current objective its effect 
is not resolvable beyond the noise. The learned strategy's gains over 
uniform cooling therefore come from its spatial structure, not from its time
variation. A constant-in-time strategy built from the
time-mean amplitudes restores temperature and the auxiliary variables
as well as the full time-varying strategy does. The absence of a
time-variation gain does not show that the seasonality is dynamically
irrelevant. The cycle is reproducible across members, but the loss is a
shallow quadratic in the amplitudes near its minimum, so the seasonal
displacement of about $1$\,K changes the segment loss by about
$0.007$\,K$^2$, below the weather noise of $0.5$\,K$^2$ per segment and
$0.04$\,K$^2$ after ten members and a year (Sec.~\ref{sec:discussion}). The tie therefore measures the flatness of
the loss, not the effect of the seasonal forcing on the climate.

\section{Prior approaches to the intervention inverse problem}\label{si:prior}

The forward problem, the climate response to a prescribed intervention,
is well-studied through coordinated model experiments
\citep{kravitz2011geomip}. The solution families for the inverse
problem make the search tractable in different ways. Optimal-pattern
studies superpose precomputed single-pattern response runs and assume
the responses add linearly \citep{banweiss2010optimization,
macmartin2013tradeoffs, brody2025optimization}. Feedback controllers
adjust a few zonal moments of the forcing with gains tuned on a
linearized design model \citep{kravitz2014explicit,
macmartin2014closedloop, kravitz2017objectives}. Inverse methods
estimate a linear response operator from Green's-function ensembles and
invert it \citep{lu2020neutral, ren2024optimal}. Reinforcement learning
treats the climate model as a black box and searches by sampling
\citep{dewitt2019sai, quan2025rl}. In feedback-controlled
multi-latitude aerosol injection, temperature objectives built from a
few zonal moments do not restore precipitation, and precipitation must
be targeted explicitly \citep{lee2020expanding}. Re-optimizing the same
aerosol-injection objectives in two versions of one climate model gives
substantially different strategies \citep{fasullo2023dependence}, so a
strategy learned in one model cannot be assumed to transfer to another.

Among gradient methods, variational data assimilation minimizes
forecast error by adjoint descent over windows of hours to days. The
ECCO ocean state estimate extends the window to decades and includes
the surface forcing fields among its adjoint control variables
\citep{forget2015ecco}, but treats the forcing as a nuisance parameter
regularized toward a prior. Conditional nonlinear optimal perturbation
finds the initial or boundary perturbation, sea-surface temperature
(SST) patterns included, that maximizes short-range error growth
\citep{mu2003cnop, song2026cnop}. Model gradients have recently been
used to optimize initial conditions for worst-case heatwave storylines
\citep{vonich2024predictability, whittaker2025storylines}, and those
optimizations end within the deterministic forecast range. In each case
the optimized quantity is a fit to observed history, a short-horizon
amplification, or an initial state, not a multi-year intervention
forcing.

\section{Area weighting of the band forcing}\label{si:areaweight}

Figs.~\ref{fig:longrun}, \ref{fig:forcing-crossmodel} and~\ref{fig:control14} show the applied cooling
weighted by the ocean area for each band with a black curve. To compute 
this net applied forcing, let $a_j$ denote the
time-mean amplitude of band $j$ in K,
$g_j(\varphi) = \exp\!\left[-(\varphi - \varphi_j)^2/2\sigma_j^2\right]$
its Gaussian profile about the center latitude $\varphi_j$
($\sigma_j = 10^{\circ}$), and $o(\varphi)$ the zonal-mean ocean
fraction. With $\cos\varphi$ weighting each latitude by its area, the thick
black line is
\begin{equation}
F(\varphi) = \sum_j a_j\, g_j(\varphi)\, o(\varphi)\cos\varphi ,
\label{eq:forcingprofile}
\end{equation}
and the thin dotted line in the same panel is the unweighted profile
$\sum_j a_j g_j(\varphi)$. Band $j$'s contribution to the global-area-mean
applied SST perturbation is
\begin{equation}
F_j = a_j A_j , \qquad
A_j = \frac{\sum_\varphi g_j(\varphi)\, o(\varphi)\cos\varphi}
           {\sum_\varphi \cos\varphi} ,
\label{eq:bandforcing}
\end{equation}
where the sum is over the grid latitudes and $A_j$ is band $j$'s effective ocean area as a fraction of
the global area, so that
$\sum_j F_j$ is the global-area-mean applied SST perturbation in K. Since
$A_j$ is fixed in time, the relation also holds at each instant,
$F_j(t) = a_j(t)\,A_j$.
Table~\ref{tab:bandforcing} lists the time-mean amplitudes, band area
fractions, and per-band forcings for the pattern-loss run of Fig.~\ref{fig:longrun}
and for the land-mean-loss run (the JAX-GCM panel of Fig.~\ref{fig:forcing-scalar}). The pattern
strategy is close to uniform in delivered forcing; its $45\degN$ band
pairs the deepest prescribed amplitude with the smallest
effective ocean area. The land-mean
strategy varies meridionally, with its equatorial band carrying 34\%
of the total applied forcing. The 14-day-horizon
pattern run of Fig.~\ref{fig:control14} has a similar spread as the
8-day run. Thus strategy is thus robust to variation in horizon length.


\begin{table}[H]
\centering
\caption{Time-mean band amplitudes and delivered forcings for the two
loss designs. $a_j$ is the time-mean prescribed amplitude of band $j$,
$A_j$ the band's effective ocean area as a fraction of global
area, and $F_j = a_j A_j$ the band's contribution to the
global-area-mean applied SST perturbation (\eqref{eq:bandforcing}).
Pattern loss, the 8-day run of Fig.~\ref{fig:longrun}; land-mean loss,
the run of the JAX-GCM panel of Fig.~\ref{fig:forcing-scalar}. Amplitudes are time means over the final 364 days of
the five-segment moving-average series.}
\label{tab:bandforcing}
\begin{tabular}{lccccc}
\toprule
 & & \multicolumn{2}{c}{Pattern loss} & \multicolumn{2}{c}{Land-mean loss} \\
\cmidrule(lr){3-4} \cmidrule(lr){5-6}
Band & $A_j$ (\%) & $a_j$ (K) & $F_j$ (K) & $a_j$ (K) & $F_j$ (K) \\
\midrule
$45\degN$   & 7.7  & $-5.42$ & $-0.42$ & $-3.69$ & $-0.29$ \\
$20\degN$   & 13.9 & $-3.55$ & $-0.49$ & $-3.57$ & $-0.50$ \\
$0^{\circ}$ & 16.8 & $-3.19$ & $-0.54$ & $-4.38$ & $-0.74$ \\
$20\degS$   & 16.1 & $-3.30$ & $-0.53$ & $-2.60$ & $-0.42$ \\
$45\degS$   & 14.3 & $-3.75$ & $-0.54$ & $-1.63$ & $-0.23$ \\
\midrule
Sum         &      &         & $-2.52$ &         & $-2.17$ \\
\bottomrule
\end{tabular}
\end{table}

\section{Control-run statistics}\label{si:controlstats}

The ten-member 8-day ensemble of Fig.~\ref{fig:longrun} (per member, 91
segments of 8 days, three-member gradient ensemble, movement penalty
$\lambda = 0.1$) has removal fractions of $92.3 \pm 0.4\%$ 
and a 2\,m temperature gain (\eqref{eq:gain}) of $G = 27.4 \pm 3.8$. 
The members of this ensemble differ in the optimizer's initial amplitudes 
and share one base weather realization; a six-run test that instead varied 
the initial condition, with the optimizer seed fixed, gave a statistically 
indistinguishable spread and ensemble-mean restoration skills agreeing to 
within 2\% for the four scored variables.
The 14-day-segment ensemble of Fig.~\ref{fig:control14}, with the
same loss and penalty, removes $94.6 \pm 1.1\%$ over three years, with $G = 22.2 \pm 2.6$.
The median in-run noise-to-signal ratio over the last ten iterations of each 
segment, for a single member, is 0.075 in the 8-day run and 0.24 in the 
14-day run.
All 91 segments of the 8-day run and 77 of 78 segments of the 14-day run 
have noise-to-signal ratio less than one.
Between the 8- and 14-day runs, four of the five time-mean band
amplitudes agree to within about 0.2\,K, the $45\degS$ amplitude
differs by 0.5\,K, and both runs place their deepest cooling on the
$45\degN$ band, where their amplitudes agree to within $0.05$\,K.
The five-member land-mean-objective ensemble of Sec.~\ref{sec:lossdesign}, otherwise identical, removes $93.2 \pm 0.3\%$ of
the land-mean warming against the pattern ensemble's $92.3 \pm 0.4\%$, with
$G = 11.9 \pm 0.5$ on 2\,m temperature against $27.4 \pm 3.8$.

\section{Replay scores and out-of-distribution warming}\label{si:replayscores}

On the $\Tm$ field the replayed strategy of Fig.~\ref{fig:replay}
scores $G = 25.3$, $2.3$, and $21.8$ in JAX-GCM, LUCIE,
and NeuralGCM, against $16.9$, $0.8$, and $10.7$ for the models' own
uniform $-4$\,K baselines, advantages of $1.5\times$ to $2.7\times$. On the
precipitation field the strategy scores $7.2$, $1.8$,
and $2.0$ against $5.2$, $1.0$, and $1.1$, advantages of
$1.4\times$ to $1.8\times$.
A 1.5-year repetition, whose 364-day averaging window is a full
seasonal cycle containing no ramp-up, reproduces the ordering in the
two hosts that completed it, with strategy-over-baseline advantages of
$1.29\times$ on temperature and $1.61\times$ on precipitation in
JAX-GCM and $2.00\times$ and $1.43\times$ in LUCIE; the
NeuralGCM run of the repetition becomes unstable after about 14 months.

Measured against each host's own unwarmed control and averaged after
the 60-day spin-up of the replay year, the flat $-4$\,K run in
JAX-GCM sits within $0.05$\,K of the target land mean,
while the learned strategy retains a warm bias of $+0.25$\,K in JAX-GCM,
$+0.04$\,K in LUCIE, and $+0.23$\,K in NeuralGCM. Thus the uniform baseline
is more effective in restoring the land-mean bias in JAX-GCM and NeuralGCM,
but is much less effective for pattern and out-of-loss variables.

Out-of-distribution behavior differs by model class. Under the uniform
$+4$\,K ocean warming, the land response
is $+3.8$\,K in JAX-GCM and $+3.9$\,K in NeuralGCM. 
LUCIE damps the land response to $+1.8$\,K,
half the warming of the other two. This is consistent
with a coordinated intercomparison of AI models under the same
prescribed $+4$\,K SST forcing, in which only a hybrid physics-AI model
reorganized regional climates in agreement with physics-based models,
while data-driven models are shown to struggle with land response to
ocean warming \citep{merchant2026zones}. 

\section{Decomposition of the temperature loss}\label{si:lossdecomp}

The two temperature terms of Eq.~\eqref{eq:loss} are the bias and
variance parts of the land-weighted mean squared error of the
segment-mean temperature. Let $T_{\mathrm{opt}}(\mathbf{x})$ and
$T_{\mathrm{tgt}}(\mathbf{x})$ be the near-surface air temperature of
the controlled run and of the target, each averaged in time over one
segment, and let $d = T_{\mathrm{opt}} - T_{\mathrm{tgt}}$. The spatial
mean $\langle f \rangle_w = \sum_{\mathbf{x}} w(\mathbf{x})\,
f(\mathbf{x})$ uses land area weights normalized to
$\sum_{\mathbf{x}} w = 1$, so $\langle \cdot \rangle_w$ is linear and
maps constants to themselves. Splitting $d$ into its land mean and the
variance about it, $d = \langle d \rangle_w + (d - \langle d
\rangle_w)$, squaring, and averaging gives
\begin{equation}
\langle d^{2} \rangle_w
\;=\;
\bigl(\langle d \rangle_w\bigr)^{2}
\;+\;
\bigl\langle \bigl(d - \langle d \rangle_w\bigr)^{2} \bigr\rangle_w ,
\label{eq:si-msedecomp}
\end{equation}
since the cross term vanishes,
$2\,\langle d \rangle_w \bigl\langle d - \langle d \rangle_w
\bigr\rangle_w = 2\,\langle d \rangle_w \bigl(\langle d \rangle_w -
\langle d \rangle_w\bigr) = 0$. The first term is the squared
land-mean bias and, by linearity,
$\langle d \rangle_w = \langle T_{\mathrm{opt}} \rangle_w - \langle
T_{\mathrm{tgt}} \rangle_w$, the squared difference of the two land
means. The second term is the spatial variance of the residual. The
land-mean objective ($\beta = 0$) therefore compares one number per
segment, the pattern objective ($\alpha = 1$, $\beta = 0.5$) keeps the
variance part at half weight, and $\alpha = \beta = 1$ recovers the
mean squared error exactly.

\clearpage
\bibliographystyle{unsrtnat}
\bibliography{refs}

@inproceedings{bonev2023sfno,
  author    = {Bonev, Boris and Kurth, Thorsten and Hundt, Christian and
               Pathak, Jaideep and Baust, Maximilian and Kashinath, Karthik
               and Anandkumar, Anima},
  title     = {Spherical {Fourier} neural operators: Learning stable
               dynamics on the sphere},
  booktitle = {Proceedings of the 40th International Conference on Machine
               Learning},
  year      = {2023},
}

@article{plotkin2019maximizing,
  title={Maximizing simulated tropical cyclone intensity with action minimization},
  author={Plotkin, David A and Webber, Robert J and O'Neill, Morgan E and Weare, Jonathan and Abbot, Dorian S},
  journal={Journal of Advances in Modeling Earth Systems},
  volume={11},
  number={4},
  pages={863--891},
  year={2019},
  publisher={Wiley Online Library}
}

@misc{dewitt2019sai,
  author = {Schroeder de Witt, Christian and Hornigold, Thomas},
  title  = {Stratospheric aerosol injection as a deep reinforcement
            learning problem},
  year   = {2019},
  note   = {arXiv:1905.07366 (ICML 2019 Climate Change AI workshop)},
}

@article{gelbrecht2023differentiable,
  author  = {Gelbrecht, Maximilian and White, Alistair and Bathiany,
             Sebastian and Boers, Niklas},
  title   = {Differentiable programming for {Earth} system modeling},
  journal = {Geoscientific Model Development},
  year    = {2023},
  volume  = {16},
  number  = {11},
  pages   = {3123--3135},
}

@article{guan2024lucie,
  author  = {Guan, Haiwen and others},
  title   = {{LUCIE}: A lightweight uncoupled climate emulator with
             long-term stability and physical consistency},
  journal = {Journal of Advances in Modeling Earth Systems},
  year    = {2025},
  volume  = {17},
  pages   = {e2024MS005152},
}

@article{hersbach2020era5,
  author  = {Hersbach, Hans and others},
  title   = {The {ERA5} global reanalysis},
  journal = {Quarterly Journal of the Royal Meteorological Society},
  year    = {2020},
  volume  = {146},
  number  = {730},
  pages   = {1999--2049},
}

@article{kochkov2024neuralgcm,
  author  = {Kochkov, Dmitrii and Yuval, Janni and Langmore, Ian and
             Norgaard, Peter and Smith, Jamie and Mooers, Griffin and
             others},
  title   = {Neural general circulation models for weather and climate},
  journal = {Nature},
  year    = {2024},
  volume  = {632},
  pages   = {1060--1066},
}

@article{kravitz2011geomip,
  author  = {Kravitz, Ben and Robock, Alan and Boucher, Olivier and
             Schmidt, Hauke and Taylor, Karl E. and Stenchikov, Georgiy
             and Schulz, Michael},
  title   = {The {Geoengineering Model Intercomparison Project (GeoMIP)}},
  journal = {Atmospheric Science Letters},
  year    = {2011},
  volume  = {12},
  number  = {2},
  pages   = {162--167},
}

@article{kravitz2014explicit,
  author  = {Kravitz, Ben and MacMartin, Douglas G. and Leedal, David T.
             and Rasch, Philip J. and Jarvis, Andrew J.},
  title   = {Explicit feedback and the management of uncertainty in
             meeting climate objectives with solar geoengineering},
  journal = {Environmental Research Letters},
  year    = {2014},
  volume  = {9},
  number  = {4},
  pages   = {044006},
}

@article{latham2012mcb,
  author  = {Latham, John and others},
  title   = {Marine cloud brightening},
  journal = {Philosophical Transactions of the Royal Society A},
  year    = {2012},
  volume  = {370},
  pages   = {4217--4262},
}

@article{crutzen2006albedo,
  author  = {Crutzen, Paul J.},
  title   = {Albedo enhancement by stratospheric sulfur injections:
             A contribution to resolve a policy dilemma?},
  journal = {Climatic Change},
  year    = {2006},
  volume  = {77},
  pages   = {211--220},
}

@article{lea2000sensitivity,
  author  = {Lea, Daniel J. and Allen, Myles R. and Haine, Thomas W. N.},
  title   = {Sensitivity analysis of the climate of a chaotic system},
  journal = {Tellus A},
  year    = {2000},
  volume  = {52},
  number  = {5},
  pages   = {523--532},
}

@article{macmartin2014closedloop,
  author  = {MacMartin, Douglas G. and Kravitz, Ben and Keith, David W.
             and Jarvis, Andrew},
  title   = {Dynamics of the coupled human--climate system resulting from
             closed-loop control of solar geoengineering},
  journal = {Climate Dynamics},
  year    = {2014},
  volume  = {43},
  pages   = {243--258},
}

@misc{metz2021gradients,
  author = {Metz, Luke and Freeman, C. Daniel and Schoenholz, Samuel S.
            and Kachman, Tal},
  title  = {Gradients are not all you need},
  year   = {2021},
  note   = {arXiv:2111.05803},
}

@article{molteni2003speedy,
  author  = {Molteni, Franco},
  title   = {Atmospheric simulations using a {GCM} with simplified
             physical parametrizations. {I}: Model climatology and
             variability in multi-decadal experiments},
  journal = {Climate Dynamics},
  year    = {2003},
  volume  = {20},
  pages   = {175--191},
}

@article{quan2025rl,
  author  = {Quan, Heng and Koll, Daniel D. B. and Lutsko, Nicholas and
             Yuval, Janni},
  title   = {Solar geoengineering strategies based on reinforcement
             learning},
  journal = {Journal of Geophysical Research: Atmospheres},
  year    = {2025},
  volume  = {130},
  pages   = {e2025JD044319},
}

@article{whittaker2025storylines,
  author  = {Whittaker, T. and {Di Luca}, A.},
  title   = {Constructing extreme heatwave storylines with differentiable
             climate models},
  journal = {Weather and Climate Dynamics},
  year    = {2026},
  volume  = {7},
  pages   = {393--410},
}

@article{banweiss2010optimization,
  author  = {Ban-Weiss, George A. and Caldeira, Ken},
  title   = {Geoengineering as an optimization problem},
  journal = {Environmental Research Letters},
  year    = {2010},
  volume  = {5},
  number  = {3},
  pages   = {034009},
}

@article{macmartin2013tradeoffs,
  author  = {MacMartin, Douglas G. and Keith, David W. and Kravitz, Ben and
             Caldeira, Ken},
  title   = {Management of trade-offs in geoengineering through optimal
             choice of non-uniform radiative forcing},
  journal = {Nature Climate Change},
  year    = {2013},
  volume  = {3},
  pages   = {365--368},
}

@article{kravitz2017objectives,
  author  = {Kravitz, Ben and MacMartin, Douglas G. and Mills, Michael J. and
             Richter, Jadwiga H. and Tilmes, Simone and Lamarque,
             Jean-Fran\c{c}ois and Tribbia, Joseph J. and Vitt, Francis},
  title   = {First simulations of designing stratospheric sulfate aerosol
             geoengineering to meet multiple simultaneous climate
             objectives},
  journal = {Journal of Geophysical Research: Atmospheres},
  year    = {2017},
  volume  = {122},
  pages   = {12616--12634},
}

@article{brody2025optimization,
  author  = {Brody, Ezra and Zhang, Yan and MacMartin, Douglas G. and
             Visioni, Daniele and Kravitz, Ben and Bednarz, Ewa M.},
  title   = {Using optimization tools to explore stratospheric aerosol
             injection strategies},
  journal = {Earth System Dynamics},
  year    = {2025},
  volume  = {16},
  pages   = {1325--1341},
}

@article{lee2020expanding,
  author  = {Lee, Walker R. and MacMartin, Douglas G. and Visioni, Daniele
             and Kravitz, Ben},
  title   = {Expanding the design space of stratospheric aerosol
             geoengineering to include precipitation-based objectives and
             explore trade-offs},
  journal = {Earth System Dynamics},
  year    = {2020},
  volume  = {11},
  pages   = {1051--1072},
}

@article{lu2020neutral,
  author  = {Lu, Jian and Liu, Fukai and Leung, L. Ruby and others},
  title   = {Neutral modes of surface temperature and the optimal ocean
             thermal forcing for global cooling},
  journal = {npj Climate and Atmospheric Science},
  year    = {2020},
  volume  = {3},
  pages   = {9},
}

@article{ren2024optimal,
  author  = {Ren, H. and Lu, J. and Hou, Z. J. and Chen, T.-C. and Leung,
             L. R. and Liu, F.},
  title   = {Neural networks to find the optimal forcing for offsetting the
             anthropogenic climate change effects},
  journal = {Artificial Intelligence for the Earth Systems},
  year    = {2024},
  volume  = {3},
  number  = {3},
}

@article{talagrand1987adjoint,
  author  = {Talagrand, Olivier and Courtier, Philippe},
  title   = {Variational assimilation of meteorological observations with
             the adjoint vorticity equation. {I}: Theory},
  journal = {Quarterly Journal of the Royal Meteorological Society},
  year    = {1987},
  volume  = {113},
  pages   = {1311--1328},
}

@article{rabier2000ecmwf,
  author  = {Rabier, Florence and J{\"a}rvinen, Heikki and Klinker, Ernst
             and Mahfouf, Jean-Fran\c{c}ois and Simmons, Adrian},
  title   = {The {ECMWF} operational implementation of four-dimensional
             variational assimilation. {I}: Experimental results with
             simplified physics},
  journal = {Quarterly Journal of the Royal Meteorological Society},
  year    = {2000},
  volume  = {126},
  pages   = {1143--1170},
}

@article{forget2015ecco,
  author  = {Forget, Ga\"el and Campin, Jean-Michel and Heimbach, Patrick and
             Hill, Chris N. and Ponte, Rui M. and Wunsch, Carl},
  title   = {{ECCO} version 4: an integrated framework for non-linear
             inverse modeling and global ocean state estimation},
  journal = {Geoscientific Model Development},
  year    = {2015},
  volume  = {8},
  pages   = {3071--3104},
}

@article{mu2003cnop,
  author  = {Mu, Mu and Duan, Wansuo and Wang, Bin},
  title   = {Conditional nonlinear optimal perturbation and its
             applications},
  journal = {Nonlinear Processes in Geophysics},
  year    = {2003},
  volume  = {10},
  pages   = {493--501},
}

@article{song2026cnop,
  author  = {Song, X. and Mu, M. and Zhan, R. and Gao, Y.},
  title   = {{SST} sensitivity of rapid intensification in {Typhoon
             Nanmadol} (2022) revealed by {CNOP}},
  journal = {Journal of Geophysical Research: Atmospheres},
  year    = {2026},
}

@article{vonich2024predictability,
  author  = {Vonich, P. T. and Hakim, G. J.},
  title   = {Predictability limit of the 2021 {Pacific Northwest} heatwave
             from deep-learning sensitivity analysis},
  journal = {Geophysical Research Letters},
  year    = {2024},
  volume  = {51},
  pages   = {e2024GL110651},
}

@misc{womack2026scenario,
  author = {Womack, Christopher B. and Bouabid, Shahine and Sokolov, Andrei
            and Salunke, Popat and Flierl, Glenn and Eastham, Sebastian D.
            and Selin, Noelle E.},
  title  = {Optimal scenario design for climate emulation},
  year   = {2026},
  note   = {arXiv:2606.19302},
}

@article{bewley2001predictive,
  author  = {Bewley, Thomas R. and Moin, Parviz and Temam, Roger},
  title   = {{DNS}-based predictive control of turbulence: an optimal
             benchmark for feedback algorithms},
  journal = {Journal of Fluid Mechanics},
  year    = {2001},
  volume  = {447},
  pages   = {179--225},
}

@article{davenport2026jcm,
  author  = {Davenport, E. H. and Madan, J. V. and others},
  title   = {{JCM} v1.1: a differentiable, intermediate-complexity
             atmospheric model},
  journal = {Geoscientific Model Development},
  year    = {2026},
  volume  = {19},
  pages   = {6451--6466},
}

@article{bala2008hydrological,
  author  = {Bala, Govindasamy and Duffy, Philip B. and Taylor, Karl E.},
  title   = {Impact of geoengineering schemes on the global hydrological
             cycle},
  journal = {Proceedings of the National Academy of Sciences},
  year    = {2008},
  volume  = {105},
  pages   = {7664--7669},
}

@article{ricke2023hydrological,
  author  = {Ricke, Katharine and Wan, Jessica S. and Saenger, Marissa and
             Lutsko, Nicholas J.},
  title   = {Hydrological consequences of solar geoengineering},
  journal = {Annual Review of Earth and Planetary Sciences},
  year    = {2023},
  volume  = {51},
  pages   = {447--470},
}

@article{fasullo2023dependence,
  author  = {Fasullo, John T. and Richter, Jadwiga H.},
  title   = {Dependence of strategic solar climate intervention on
             background scenario and model physics},
  journal = {Atmospheric Chemistry and Physics},
  year    = {2023},
  volume  = {23},
  pages   = {163--182},
}

@misc{merchant2026zones,
  author = {Merchant, Charlotte C. and Kl{\"o}wer, Milan and
            Stanley-Clamp, Bradley and H{\"o}ver, Maren and
            Michel, Simon L. L. and Groot, Edward and
            Christensen, Hannah M.},
  title  = {How do {AI} climate models respond to warming across
            climate zones?},
  year   = {2026},
  note   = {arXiv:2608.17986},
}

@article{guan2025lucie,
  title={LUCIE-3D: A three-dimensional climate emulator for forced responses},
  author={Guan, Haiwen and Arcomano, Troy and Chattopadhyay, Ashesh and Maulik, Romit},
  journal={arXiv preprint arXiv:2509.02061},
  year={2025}
}

@article{hassanzadeh2016lrf,
  author  = {Hassanzadeh, Pedram and Kuang, Zhiming},
  title   = {The linear response function of an idealized atmosphere.
             Part {I}: Construction using {G}reen's functions and
             applications},
  journal = {Journal of the Atmospheric Sciences},
  year    = {2016},
  volume  = {73},
  number  = {9},
  pages   = {3423--3439},
  doi     = {10.1175/JAS-D-15-0338.1},
}

\end{document}